\documentclass[11pt]{article}
\usepackage{jheppubmod}
\pdfoutput=1
\usepackage{setspace}
\usepackage{amsmath}
\usepackage{amsfonts}
\usepackage{amssymb,amsthm}
\usepackage{graphicx}
\usepackage{multirow}
\usepackage{float}
\usepackage{tikz}
\usetikzlibrary{shapes,arrows,shadings,shadows}
\usepackage{graphicx}
\usepackage[compat=1.1.0]{tikz-feynman} 
\usepackage{hyperref}
\usepackage[english]{babel}
\usepackage{subcaption}
\usepackage[justification=centering,singlelinecheck=false]{subcaption}
\usepackage[justification=RaggedRight,singlelinecheck=false]{caption}
\usepackage{amsfonts}
\usepackage{braket}
\usepackage{mathrsfs}
\usepackage{empheq}
\usepackage{lipsum}
\usepackage[shortlabels]{enumitem}
\usepackage{pgfplots}
\usepackage{bigints}
\usepackage{tikz, pgf}
\usepackage{amsmath}
\usepackage{tikz}
\usetikzlibrary{positioning,arrows.meta}
\usepackage{afterpage}
\usetikzlibrary{decorations.pathmorphing}
\usetikzlibrary{arrows.meta}
\tikzset{%
  >={Latex[width=0.2mm,length=2mm]},
            base/.style = {rectangle, rounded corners, draw=black,
                           minimum width=4cm, minimum height=1cm,
                           text centered, font=\sffamily},
}
\usetikzlibrary{calc,bending,decorations.markings}
\usepackage{setspace}
\usepackage{multirow}
\usepackage{graphicx}
\usepackage{dcolumn}
\usepackage{bm}
\usepackage{xcolor} 

\begin{document}
\title{Hunting long-lived doubly charged scalars at the HL-LHC}
\author[1]{Biplob Bhattacherjee}
\author[2]{Rituparna Ghosh}
\author[2]{Swagata Mukherjee}
\author[2]{Nabin Kumar Pidikaka}

\affiliation[1]{Centre for High Energy Physics, Indian Institute of Science, Bangalore 560012, India}
\affiliation[2]{Department of Physics, Indian Institute of Technology, Kanpur 208016, India}

\emailAdd{biplob@iisc.ac.in}
\emailAdd{rituparnag@iitk.ac.in}
\emailAdd{swagata@iitk.ac.in}
\emailAdd{nabinkp24@iitk.ac.in}
\vspace{0.5cm}

\abstract{ This work studies the collider phenomenology of long-lived doubly charged scalars. After reviewing the existing searches for the doubly charged scalar in both the prompt and the long-lived regimes, we identify an intermediate range of proper decay length, namely, $\mathcal{O}(0.1~\mathrm{mm}) \lesssim c\tau \lesssim \mathcal{O}(100~\mathrm{mm})$, where conventional searches lose sensitivity. We investigate the prospects for probing the doubly charged scalar both in the presence and in the absence of the $\Delta L=2$ Yukawa coupling of the $SU(2)_{L}$ complex triplet. In the presence of this coupling, $H^{\pm\pm}$ can be long-lived only for masses in the range $100$-$150~\mathrm{GeV}$, whereas in the fermiophobic scenario (i.e., in the absence of the $\Delta L=2$ coupling), the range extends to $\mathrm{TeV}$ scale. We propose a displaced-vertex search at the HL-LHC for doubly charged scalars with masses up to $\approx 1~\mathrm{TeV}$ and $c\tau=10$-$100~\mathrm{mm}$ in the fermiophobic scenario, while a benchmark point with a doubly charged scalar of mass $120~\mathrm{GeV}$ and $c\tau = 5~\mathrm{mm}$ is considered when the complex triplet scalar couples to leptons. We show that a cut on the invariant mass of the displaced vertex as reconstructed from the associated tracks can strongly suppress Standard Model backgrounds. We present projected limits on the Drell-Yan pair-production cross section for the doubly charged scalar at $\sqrt{s}=14~\mathrm{TeV}$ with an integrated luminosity of $3000~\mathrm{fb}^{-1}$, considering two illustrative assumptions for the residual background. Displaced-vertex searches thus probe a region complementary to those covered by prompt and heavy stable charged-particle searches.
}

\maketitle

\newpage
\section{Introduction}

Various observations such as the baryon asymmetry of the Universe (BAU)~\cite{wmap,wmap1,wmap2,wmap3,wmap4,Pl1,Pl2}, and
the nonzero masses of neutrinos~\cite{superk1998,Kayser2012,Kayserqm}, have pointed to the existence of physics beyond the Standard Model (BSM). Over the past several decades, particle physics experiments such as the Large Electron-Positron Collider (LEP) and the Tevatron, followed by the Large Hadron Collider (LHC), have extensively searched for signatures of new physics, but despite these efforts, no conclusive evidence for BSM particles has yet been observed.

The absence of new physics in conventional searches has motivated increasing interest in unconventional signatures of BSM physics. Most collider searches assume that BSM particles, after being produced, decay promptly into Standard Model (SM) particles. The lack of evidence in favor of new physics in such searches suggests that, even if BSM particles exist at the electroweak scale, their interactions with the SM might be sufficiently weak. Such suppressed couplings naturally lead to smaller decay widths and consequently longer lifetimes, allowing the production and decay vertices of these particles to be spatially separated within the detector. Since the efficiency of prompt searches deteriorates rapidly as the decay length increases, long-lived particles (LLPs) require dedicated search strategies that exploit displaced signatures rather than prompt decay products.

This work will focus on the exploration of the $SU(2)_{L}$ triplet scalar extension of the SM via the search for doubly charged scalars in the small-decay-width regime, which, as we show below, has received comparatively little attention. In many extensions of the SM that aim to explain the generation of neutrino masses or BAU, the scalar sector is extended by adding an $SU(2)_L$ triplet scalar~\cite{seesaw1,seesaw2,trip1,trip2,baut1,baut2}.   Moreover, such a scalar sector also appears in composite Higgs models that aim to solve the naturalness problem or explain the electroweak symmetry breaking~\cite{gm1,gm2,ch1,ch2}. In models which simultaneously address neutrino mass generation, dark matter, and the SM fermion mass hierarchy, the $H^{\pm\pm}$ frequently acts as a portal to vector-like fermions (VLFs), altering the scalar sector's landscape \cite{ Bahrami:2013bsa, Bhattacharya:2018fus, Ghosh:2018drw}. While the explanation of neutrino mass or BAU strongly relies upon the coupling of the triplet scalar to leptons, the second scenario survives even if the triplet scalar turns out to be fermiophobic. Here we will be completely agnostic about the UV completion of the theory and will only concentrate on the collider search for such an extension of the SM comprehensively. To date, searches for triplet scalars have been focused primarily on the doubly charged scalar, assuming it decays promptly into either same-sign dileptons~\cite{OPAL_LEP, ATLAS_2014, atlas_dilep2, ATLAS_2022} or same-sign $W$ boson pair~\cite{ww1,atlas_2021,ww2}. Depending on the underlying theoretical scenario, specifically, whether or not a lepton-number-violating Yukawa coupling $(y_{\ell\ell})$ of the triplet scalar is present, the region of parameter space accessible through such prompt signatures can vary significantly.

For instance, a doubly charged scalar with a mass of $200~\mathrm{GeV}$ undergoes a prompt decay into same-sign 
$W$ bosons when the triplet vacuum expectation value~(vev), $v_\chi \gtrsim 10^{-4}~\mathrm{GeV}$, and exhibits a delayed decay into the same final state for smaller vev in the absence of lepton-number-violating Yukawa couplings~\cite{gmllp}. In contrast, when such Yukawa coupling is present, the same particle decays promptly into both same-sign diboson and same-sign dilepton final states over a vast region of parameter space. Therefore, a complete experimental exploration of triplet scalar models requires dedicated searches for the long-lived particle signatures predicted in these scenarios.

Here we have performed a detailed collider analysis in the context of the High-Luminosity Large Hadron Collider~(HL-LHC), focusing on the region of parameter space where the predicted particles, primarily the triplet-dominated ones, undergo a delayed decay. We have considered the theoretical scenario, namely the 
Georgi-Machacek~(GM)~\cite{gm1,cg} model for the simulation of signals here. We consider the lepton number-violating Yukawa coupling of the complex triplet, which is usually ignored in the context of the GM model. Although such an extension of the SM is well known for its 
ability to accommodate a high triplet vev consistently with the electroweak precision tests, in this work we will scan over the full range of the triplet vev and explore the potential to probe different regions of triplet vev at the collider. The model is used here solely for simulation purposes, and the study has been kept as model-independent as possible. No model parameter other than the triplet vev and the mass of the doubly charged scalar is relevant for the study.

The paper is organized as follows: Sec.~\ref{model} discusses the theoretical scenario and marks the region of parameter space 
relevant to this study. In sec.~\ref{constraint}, we discuss the constraints on the mass of the doubly charged scalar. Here we have focused on the constraints that depend only on the mass of the doubly charged scalar and the triplet vev. 
Sec.~\ref{col} describes the whole simulation and the analysis procedure followed. In sec.~\ref{result}, 
the results are presented, and we finally conclude in sec.~\ref{conc}.

\section {Theoretical scenario}
\label{model}
We have considered the Georgi-Machacek model~\cite{gm1,cg} comprising an $SU(2)_{L}$ complex triplet $\chi$ and real triplet $\xi$ along with the Standard Model $SU(2)_{L}$ doublet $\phi$, for its ability to accommodate a vast range of triplet vacuum expectation values ranging from a few $\mathrm{meV}$ to about a hundred $\mathrm{GeV}$~\cite{decoup, ghosh}. It is well-known that by the manifestation of a custodial $SU(2)$ symmetry, this model can accommodate a high triplet vev consistently with oblique parameters and in this high triplet vev regime the lepton number violating Yukawa coupling of the $SU(2)$ triplet does not play any significant role as long as one generates the neutrino masses via type-II seesaw mechanism~\cite{t2-1,t2-2,t2-3,t2-4,t2-5,t2-6}. Even if one starts with a Lagrangian where the $\Delta L =2$ Yukawa coupling of a complex triplet scalar is set to zero, the higher-order effects will introduce such a coupling, and the coupling strength will depend on various model parameters. Hence, in this work we will consider both scenarios: (a) $y_{\ell\ell}$ playing a non-trivial role; (b) $y_{\ell\ell}$ is negligible.

The most relevant parts of the Lagrangian for this study are the kinetic term involving the $SU(2)_{L}$ triplets $(\mathcal{L}^{T}_{\text{kin}})$\cite{gm1} and the Yukawa coupling of the $SU(2)_{L}$ complex triplet $(\mathcal{L}_{Y})$ 
 
\begin{equation}
\mathcal{L}^{T}_{\rm kin} =  
\frac{1}{2}\,\mathrm{Tr}\!\left[(D_\mu X)^\dagger (D^\mu X)\right],
\end{equation}
where 
\begin{equation}
X = 
\begin{pmatrix} 
\chi^{0} & \xi^+ & \chi^{++} \\ 
\chi^{-} & \xi^0 & \chi^+ \\ 
\chi^{--} & -\xi^{+*} & \chi^{0\star} 
\end{pmatrix}.
\end{equation}
with covariant derivative,
\begin{equation}
D_\mu X = \partial_\mu X 
+ i g W_\mu^a T^a X 
- i g' B_\mu X T^3,
\end{equation}
where $T^a$ ($a=1,2,3$) are the $SU(2)$ generators in the 
triplet (adjoint) representation.

The $\Delta L =2$ Yukawa coupling of the complex triplet is given by,

\begin{align}
\mathcal{L}_Y
=&
-\frac{1}{2}Y_{ij}
\Bigg[
\nu_i^{T}C\nu_j\,\chi^0
-\frac{1}{\sqrt{2}}
\left(
\nu_i^{T}C\ell_j
+\ell_i^{T}C\nu_j
\right)\chi^+
-\ell_i^{T}C\ell_j\,\chi^{++}
\Bigg]
+\mathrm{h.c.}.
\end{align}

\begin{figure}[htb!]
     \begin{subfigure}[b]{0.5\textwidth}
         \centering
         \includegraphics[width=\textwidth]{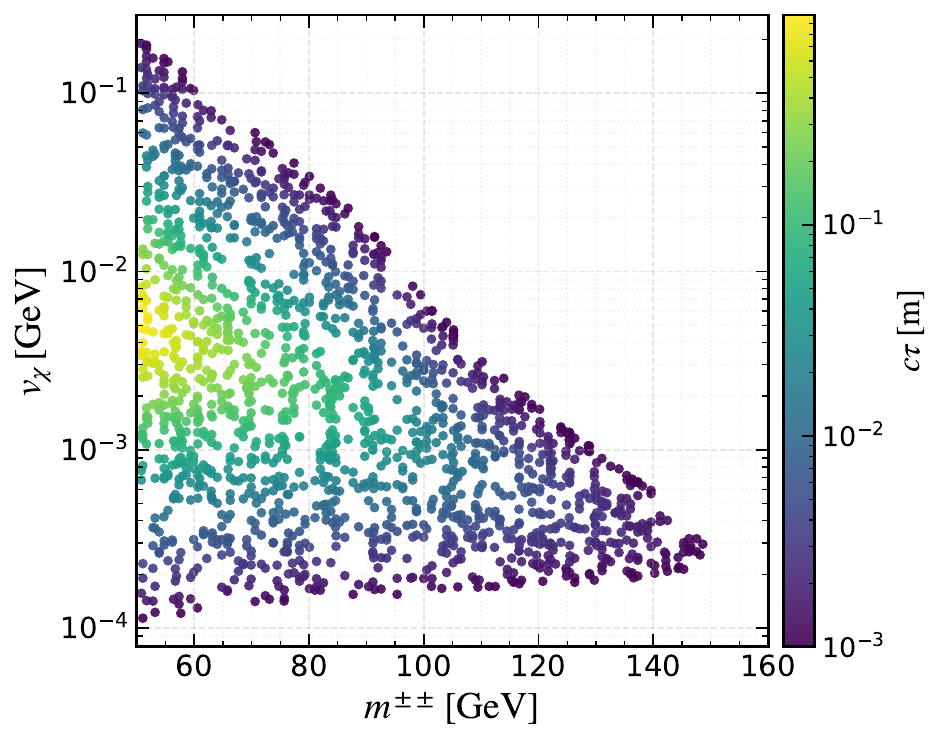}
         \caption{}
         \label{llp_yuk_reg}
     \end{subfigure}
     \hfill
     \begin{subfigure}[b]{0.5\textwidth}
         \centering
         \includegraphics[width=\textwidth]{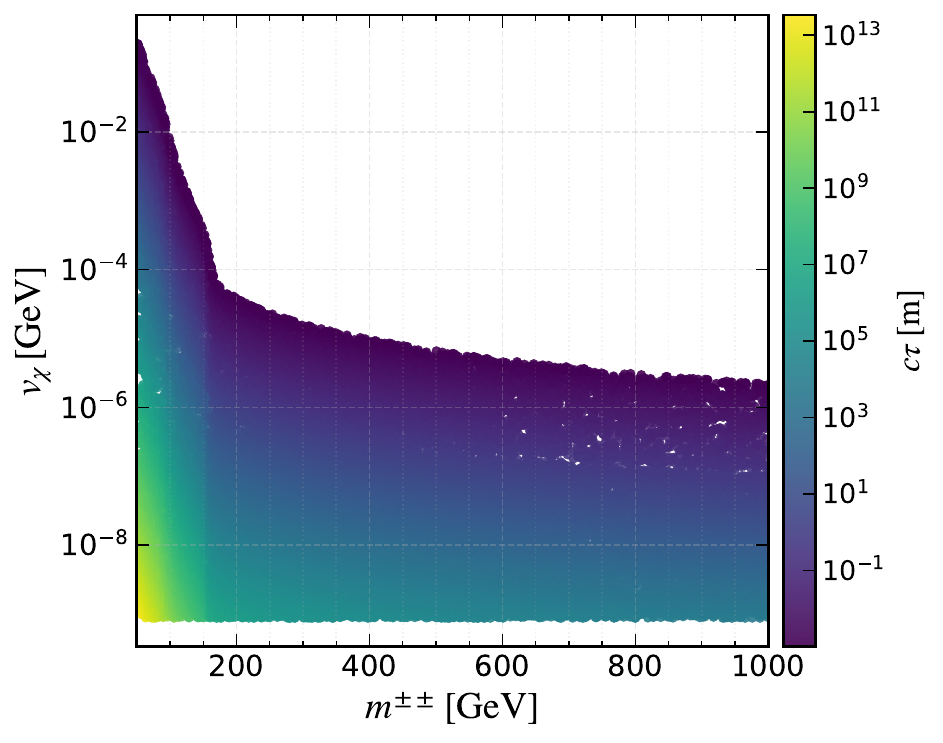}
         \caption{}
         \label{llp_noyuk_reg}
     \end{subfigure}
     \caption{Left: Region of parameter space where the doubly charged scalar becomes long-lived in the presence of Yukawa coupling of the triplet. Right: Region of parameter space where the doubly charged scalar becomes long-lived in the absence of triplet Yukawa coupling.}
     \label{regplot}
    \end{figure}

\begin{itemize}
    \item {$y_{\ell\ell} \ne 0$}

The Yukawa coupling introduces neutrino masses for the neutrinos as well as mixing among them. For simplicity, we have considered the $Y_{ij}$ to be diagonal and hence the mass of the $i^{th}$ mass eigenstate of the neutrinos is given by,

\begin{equation}
\label{mnu}
(m_\nu)_{i} = v_\chi Y_{ii}.
\end{equation}   

 It is well-known that, depending on the hierarchy of the neutrino masses, two distinct mass patterns appear: (1)~the normal hierarchy: $m_3>m_2>m_1$, and (2)~the inverted hierarchy: $m_2>m_1>m_3$, where $m_i$ is the mass of the $i^{\text{th}}$ mass eigenstate of the neutrino mixing matrix. Under our simplistic assumption of the Yukawa coupling being diagonal, the flavor and mass eigenstates coincide, and the normal and the inverted hierarchy imply $Y_{33} > Y_{22} > Y_{11}$ and $Y_{22} > Y_{11} > Y_{33}$, respectively. 
 
 The benchmark in this scenario has been chosen such that the constraint on the sum of the neutrino masses, 
 i.e., $\Sigma_i \, m_{\nu_i} \le 0.1~ \text{eV}$\cite{Pl2}, is satisfied. For a given $v_\chi$ the Yukawa couplings were determined in such a way that the constraint on the difference of squared masses as obtained from the atmospheric and solar neutrino data is satisfied, i.e., $\Delta m_{21}^2 \simeq 7.5\times 10^{-5}~ \text{eV$^{2}$}$ and  $|\Delta m_{31}^2| \simeq 2.4\times 10^{-3}~ \text{eV$^{2}$}$ is preserved, where $\Delta m_{ij}^2 = m_{i}^2 - m_{j}^2$, with $m_{i}$ and $m_{j}$ being the mass of the neutrinos~\cite{nufit}.

In this scenario, the doubly charged scalar decays both into same-sign $W$ bosons and same-sign dileptons. The coupling of $H^{\pm\pm}$ to $W^\pm W^\pm$ is given by,
\begin{equation}
\label{ghww}
    g_{H^{\pm\pm}W^\mp W^\mp} = \frac{2\,g\, m_W \, v_\chi}{v}
\end{equation}
where $g$ is the $SU(2)_L$ gauge coupling, $m_W$ is the mass of the $W$ boson and $v$ is the standard electroweak vev.

As evident from equation~\eqref{mnu}, the Yukawa coupling is inversely proportional to the triplet vev, and as one lowers the triplet vev, the partial decay width corresponding to the $WW$ channel decreases, but the leptonic decay width goes up. Hence, in this scenario the doubly charged scalar appears to be long-lived only in a range of $v_\chi$ as depicted in fig.~\ref{llp_yuk_reg}. On the mass scale of the doubly charged scalar, this range corresponds to the mass of the doubly charged scalar staying below the $WW$ threshold, as shown 
in fig.~\ref{llp_yuk_reg}.

\item $y_{\ell\ell} =0$

In this scenario, the benchmark has been chosen such that the doubly charged scalar decays exclusively into same-sign $WW$ bosons. Since the $H^{\pm\pm} W^{\mp} W^{\mp}$ coupling depends only on the triplet vev, as is clear from equation \eqref{ghww}, lowering $v_\chi$ monotonically decreases the decay width of the doubly charged scalar. Hence, unlike the previous scenario, in this case, as the triplet vev goes below $10^{-4}~\mathrm{GeV}$, $H^{\pm\pm}$ starts emerging as an LLP, as shown in fig.~\ref{llp_noyuk_reg}. Below the $WW$ threshold, $H^{\pm\pm}$ emerges as an LLP for a triplet vev as high as $10^{-2}~\mathrm{GeV}$.
\end{itemize}

From the discussion we have just presented above, it is clear that the important parameters for this study are the triplet vev and the mass of the doubly charged scalar. Hence, a detailed discussion of the model is irrelevant here, and we will discuss only those experimental constraints with regard to this model that depend only on these two parameters. 

The parameter space shown in fig.~\ref{regplot} respects the constraints coming from unitarity and vacuum stability while scanning the parameter space~\cite{decoup, uni}. The lowest limit on the mass of the doubly charged scalar comes from the search for $H^{\pm\pm}$ in $Z$ boson decay, which puts a limit of $\approx 45.6~\mathrm{GeV}$~\cite{lowest_lim}. Though this search assumes the doubly charged Higgs exclusively decays into same-sign dileptons, we have kept the lowest limit for our scan to be $50~ \mathrm{GeV}$. While for the $y_{\ell\ell} \ne 0$ scenario, the region of parameter space with the doubly charged scalar behaving as an LLP remains below the $WW$ threshold, i.e., 
between $50-150~ \mathrm{GeV}$; for the second scenario where the triplet couples to gauge bosons only, it can range up to a very high mass. We have deliberately truncated the scan for fig.~\ref{regplot} at $1~\mathrm{TeV}$ because, above that scale, the possibility of detecting such an exotic scalar produced by Drell-Yan~(DY) pair production fades in the context of the LHC. Only for the first scenario, the flavor-changing processes become important, but they start to dominate below $10^{-6}~\mathrm{GeV}$ of triplet vev~\cite{trip2}. We have not considered any constraint that depends on any model parameter other than the mass of the doubly charged scalar and the triplet vev, as these are the only parameters of interest in this study.


\section{Constraint from the existing data}
\label{constraint}

The doubly charged scalar has been extensively searched for both at the LEP and the LHC under different assumptions for the decay lengths. Here we shall give a brief description of the constraints imposed by the experimental searches for different lifetimes of the doubly charged scalar and also point out the region in the mass vs lifetime plane that needs to be searched with dedicated efforts.

\subsection{Constraint from prompt searches}

Since the decay profile of an unstable particle follows an exponential distribution, a fraction of signal events may still satisfy the primary vertex (PV) reconstruction requirements employed in experimental analyses, even when the particle is long-lived. Consequently, regions of parameter space corresponding to longer decay lengths may still be constrained by searches designed for promptly decaying particles.

In the case of doubly charged scalars, the most stringent constraints arise from searches in the same-sign dilepton final state and same-sign diboson final states. 
\begin{itemize}
    \item Searches for doubly charged scalars decaying into same-sign dileptons have been carried out since the LEP era. LEP excluded doubly charged scalar masses up to 100~GeV~\cite{OPAL_LEP}, while the LHC subsequently improved the exclusion limit to $550~\mathrm{GeV}$ during Run-1~\cite{ATLAS_2014} and to 1.1~TeV during Run-2~\cite{ATLAS_2022}.
    
    \item The ATLAS experiment has searched for doubly charged scalars produced via the DY process and decaying into same-sign $W$ boson pairs, excluding masses in the range of $200-220~\mathrm{GeV}$ using 36.1~$\text{fb}^{-1}$ of data~\cite{ww1}. A subsequent analysis extended the excluded mass range to $200-350~\mathrm{GeV}$~\cite{atlas_2021}.
\end{itemize} 

\begin{figure}[htbp!]
	\centering
    \subfloat[]{\label{prmpt_eff}\includegraphics[width=0.35\textwidth]{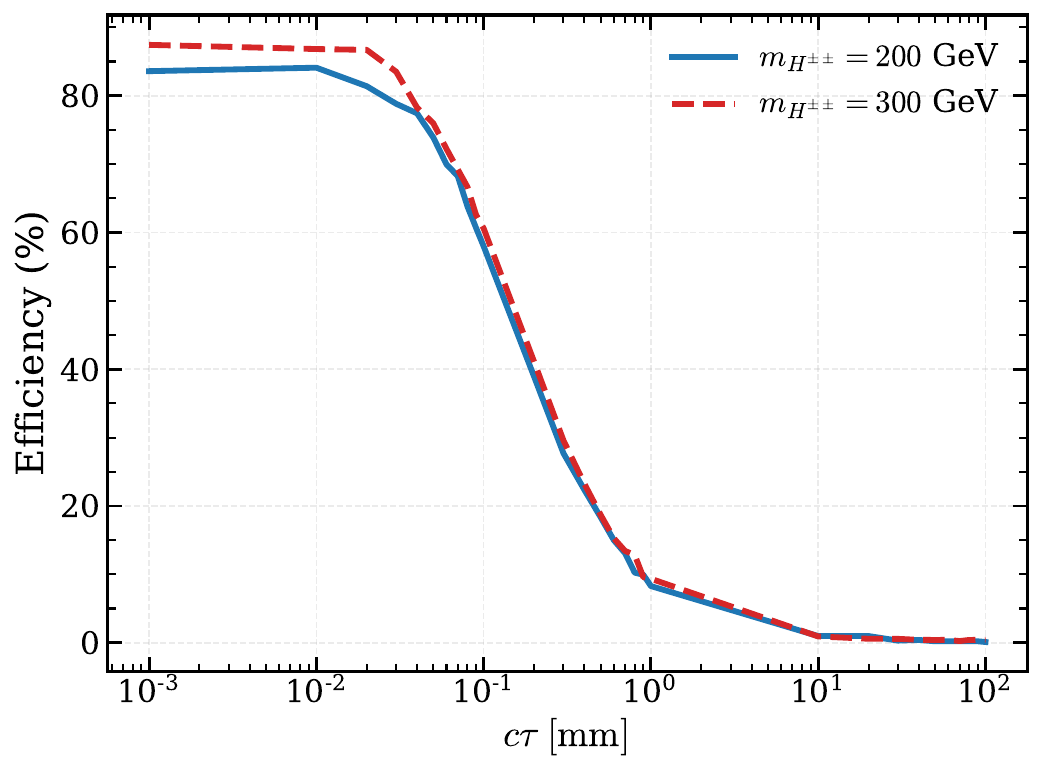}}
    \subfloat[]{\label{trk_eff}\includegraphics[width=0.35\textwidth]{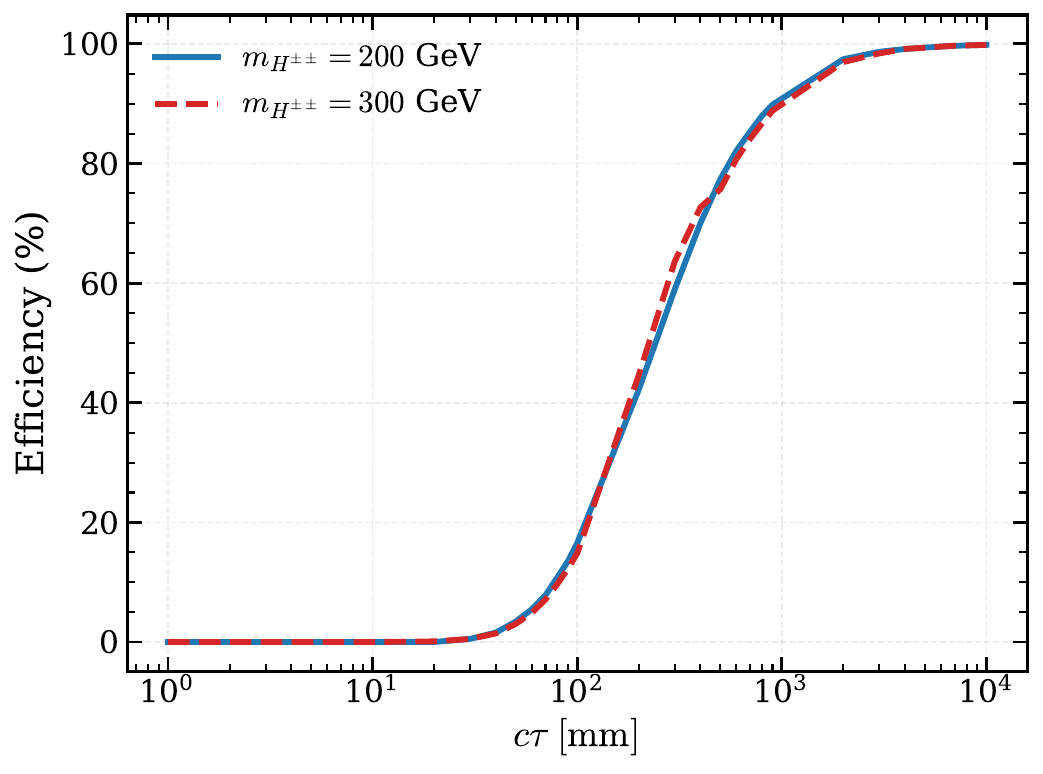}}
	\subfloat[]{\label{hscp_eff}\includegraphics[width=0.35\textwidth]{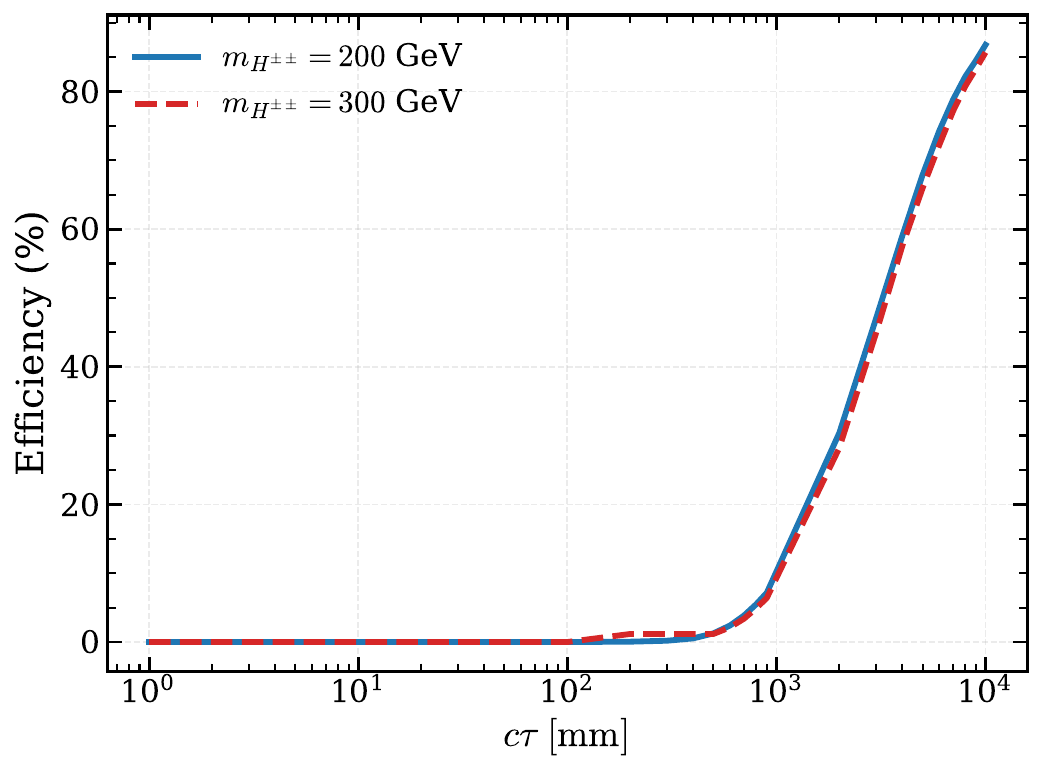}}
    \caption{Left: Efficiency of primary vertex reconstruction criteria, for prompt searches at $\sqrt{s}=13$~TeV $pp$ collisions at the LHC for different proper decay lengths; Middle: Efficiency of $L_{xy} > 30~\mathrm{cm}$ criterion where \texttt{tracker-only} analysis works;  Right: Efficiency of the requirement of $L_{xy} > 4~\mathrm{m}$ where the \texttt{track+ToF} analysis works. } 
	\label{llp_const}
\end{figure}

This brings us to the question of the validity of these limits when the doubly charged scalar acquires a measurable lifetime. For instance, the analysis in ref.~\cite{atlas_2021} requires the presence of at least two leptons having tracks that satisfy $\text{Sig}[d_0] < 3$ and $z_{0}\sin\theta < 0.5~\text{mm}$ for muons, and $\text{Sig}[d_0] < 5$ and $z_{0}\sin\theta < 0.5~\text{mm}$ for electrons, with $d_0$ being the transverse impact parameter, $z_0$ being the distance of the closest approach point from the PV along beam-line, and $\theta$ being the angle between beam-line and the direction from PV to the closest approach point.
To investigate the impact of a finite decay length, we simulated pair production of doubly charged scalars via the DY process, followed by their decays into same-sign $W$ boson pairs, with the $W$ bosons subsequently decaying leptonically. Fig.~\ref{prmpt_eff} shows the efficiency of satisfying the PV reconstruction criteria as a function of the scalar decay length. As is evident from the figure, the efficiency decreases rapidly once the decay length reaches $\mathcal{O}(\mathrm{mm})$ or larger. Consequently, prompt searches become insensitive in this regime, rendering the corresponding exclusion limits inapplicable.

\subsection{Constraint from direct search for long-lived particles}

Since the limits coming from prompt searches cannot be applied to particles having a decay length of $\mathcal{O}$(mm), collider experiments have specifically searched for particles with longer decay lengths since the LEP era. Here we will discuss the searches for long-lived doubly charged scalars briefly. Direct searches for charged long-lived particles have mainly relied upon three strategies: signatures involving disappearing tracks, time-of-flight measurements, and ionization energy loss measurements.

\subsubsection{Search for heavy stable charged particles} 

Both LEP~\cite{DELPHI_HSCP} and the Tevatron~\cite{Teva_hscp} have searched for long-lived doubly charged scalars as candidates for heavy stable charged particles (HSCPs). Consequently, these analyses require the doubly charged scalar to traverse the detector as a charged track and be reconstructed as a global muon with a high ionization energy loss ($dE/dx$) in the tracker that is significantly different from that of SM particles. Owing to the relatively low center-of-mass energies of these experiments, these searches are sensitive to doubly charged scalar masses only when the scalar has a decay length of approximately $\mathcal{O}(\mathrm{m})$.

References~\cite{aad2011search,atlas_2016_1,atlas_2016_2,atlas_2019,atlas_dedx_2022} present searches by the ATLAS Collaboration for heavy stable charged particles using anomalously large ionization energy loss ($dE/dx$) and \texttt{time-of-flight (ToF)} measurements in 
Run-1~\cite{aad2011search} and in 
Run-2~\cite{atlas_2016_1,atlas_2016_2,atlas_2019,atlas_dedx_2022}. These analyses are interpreted within supersymmetric scenarios containing long-lived gluinos, charginos, sleptons, stops, or R-hadrons. The searches assume that a charged LLP traverses the inner detector and calorimeters before reaching the muon spectrometer either as a charged or as a neutral particle. Accordingly, the analyses have been performed using two complementary approaches: a \texttt{tracker-only} analysis and a \texttt{track+ToF} analysis. The former is employed in refs.~\cite{atlas_2016_1,atlas_2016_2,atlas_2019}, whereas refs.~\cite{atlas_dedx_2022} exploits the \texttt{track+ToF} strategy. In the \texttt{tracker-only} analysis, events are required to exhibit significant missing transverse momentum, which is used for triggering, together with a charged-particle track in the inner detector exhibiting anomalously large $dE/dx$. The \texttt{track+ToF} analysis targets events in which the charged LLP is reconstructed in the muon spectrometer, allowing its velocity to be determined from the \texttt{time-of-flight} measurement. The search for LLPs in supersymmetric contexts was performed in Run-2 as well. The analyses described in refs.~\cite{atlas_2016_1,atlas_2016_2} are dedicated to R-hadrons and applicable to R-hadron masses above $600~\mathrm{GeV}$ for lifetimes of the order of $\mathcal{O}(\mathrm{ns})$. In contrast, refs.~\cite{atlas_2019,atlas_dedx_2022} extend the interpretation to include charginos and sleptons in addition to R-hadrons. These analyses set exclusion limits on the masses and lifetimes of charginos and sleptons for masses above approximately $600~\mathrm{GeV}$, and place upper limits on the production cross sections of gluinos and R-hadrons for masses above about $1~\mathrm{TeV}$. Since the production cross section of these particles can differ substantially from that of a doubly charged scalar, whose production proceeds solely through electroweak interactions, these limits are not directly applicable to our scenario and are therefore not considered in the present study. ATLAS has also perform a search for multi charged particles as described in \cite{atlas_mcp} and the corresponding limit is directly applicable one our scenario excluding the doubly charged scalar with mass up to $\approx 1~\mathrm{TeV}$. Since this search requires the LLP to reach the muon chamber, the exclusion is applicable to particle which are almost detector stable.

Reference~\cite{cms_2016} presents a search for HSCPs by the CMS experiment using two complementary strategies: a \texttt{tracker-only} analysis and a \texttt{track+ToF} analysis, where the time of flight of the LLP to the muon system is exploited to discriminate signal from background. Unlike the ATLAS searches, the CMS analysis employs a global muon trigger in addition to the conventional missing transverse momentum trigger. The \texttt{track+ToF} analysis is sensitive to LLPs with lifetimes of $\mathcal{O}(\mathrm{ns})$ and masses above approximately $200~\mathrm{GeV}$. In contrast, the \texttt{tracker-only} analysis is capable of probing scenarios in which the LLP traverses only the inner tracking system, requiring the particle to travel a radial distance of approximately $30~\mathrm{cm}$\cite{atlas_2016_1}. This requirement arises from the demand that the reconstructed track must contain at least six hits in the silicon strip tracker. Ref.~\cite{cms_dedx_2025} updates the previous CMS HSCP search. Unlike the earlier analysis, it no longer employs a trigger based solely on the presence of a global muon and instead requires events to satisfy both a global muon trigger and a calorimeter-based missing transverse momentum trigger. Specifically, the selected events must contain at least one global muon with $p_T > 50~\mathrm{GeV}$ and missing transverse momentum exceeding $170~\mathrm{GeV}$. Consequently, this search is primarily sensitive to LLPs that are effectively detector-stable, corresponding to lifetimes of approximately $\gtrsim 1~\mathrm{ns}$. Since both analyses rely exclusively on measurements of the ionization energy loss, $dE/dx$ (and related discriminating observables in ref.~\cite{cms_dedx_2025}), together with the particle velocity $\beta$, which depend only on the mass and momentum of the LLP, the resulting constraints are largely model-independent. Furthermore, both analyses present their results as upper limits on the DY production cross section of charged LLPs. For a given LLP mass, ref.~\cite{cms_dedx_2025} provides the most stringent constraints, excluding the existence of charged LLPs with masses in the range $800$-$1400~\mathrm{GeV}$ for lifetimes of the order of $\mathcal{O}(\mathrm{ns})$. 

As it has been argued above, the \texttt{tracker-only} analysis will impose a constraint when the HSCP traverses $\approx~ 30~\mathrm{cm}$ \cite{atlas_2016_1} in the radial direction, i.e., $L_{xy} \ge 30~\mathrm{cm}$, where $L_{xy}$ is the distance traveled by the LLP  in the transverse plane to the beam direction before decaying. While for muon chamber-based analysis $L_{xy} \ge 4~\mathrm{m}$ is required since the muon chambers at CMS and ATLAS start at $4~\mathrm{m}$ and $5~\mathrm{m}$ respectively~\cite{cms-tdr,atlas-tdr}. Fig.~\ref{trk_eff} and fig.~\ref{hscp_eff} show the efficiency of these two criteria as a function of the proper decay length of the LLP, respectively.

\subsubsection{Constraint from searches for displaced vertex} 
Long-lived particles with shorter decay lengths, such that they decay before reaching the muon system, have also been searched for in experiments. These searches at LEP employed two complementary strategies: a small impact parameter search and a secondary vertex (or kink) search~\cite{DELPHI_HSCP}. In both cases, in the context of doubly charged scalars, the analyses were interpreted under the assumption that the doubly charged scalar decays exclusively into the $\tau^\pm\tau^\pm$ final state. The small impact parameter search targets LLPs decaying within the beam pipe. Events are required to contain four low-multiplicity jets, while the associated charged-particle tracks must satisfy transverse and longitudinal impact parameters smaller than $4~\mathrm{cm}$. The secondary vertex search, on the other hand, is designed to identify LLPs decaying within the tracking volume. For the DELPHI detector, this corresponds to a radial region extending from approximately $6~\mathrm{cm}$ to $1.12~\mathrm{m}$. The analysis reconstructs the LLP trajectory together with that of its charged decay product and requires the decay vertex to lie within the vertex detector, corresponding to a radial distance below approximately $11~\mathrm{cm}$. The combination of the small impact parameter and secondary vertex searches excludes doubly charged scalars with masses below approximately $100~\mathrm{GeV}$ for decay lengths up to $\mathcal{O}(10~\mathrm{cm})$, assuming that $H^{\pm\pm}$ decays predominantly into $\tau^\pm\tau^\pm$. Although this result is derived for the $\tau^\pm\tau^\pm$ decay mode, the reconstruction strategy underlying the secondary vertex search is not expected to depend strongly on the specific visible final state. It is therefore reasonable to expect comparable sensitivity for decay modes such as $H^{\pm\pm}\rightarrow e^\pm e^\pm$, $\mu^\pm\mu^\pm$, or $W^\pm W^\pm$, provided that the corresponding decay products satisfy the reconstruction requirements. Motivated by these considerations, we conservatively restrict our benchmark points to masses above $100~\mathrm{GeV}$.

The LHC has also searched for LLPs with relatively short decay lengths through displaced vertex and displaced-track signatures. In particular, the CMS Collaboration has performed searches for displaced vertices for a signal consisting of  pair-production of LLPs and their subsequent decay into quark-antiquark 
pairs~\cite{CMS_disp_jet_2019, CMS_disp_jet_2020}. Owing to their largely model-independent event selection, these analyses are among the most relevant existing searches for our study in the regime of the small decay length. Both analyses exploit the track multiplicity associated with the displaced vertex as one of the primary discriminating observables. However, they are specifically designed for neutral LLPs, whose displaced vertices exhibit track multiplicities that can differ significantly from those originating from a charged LLP. Another important observable is the cluster RMS, which incorporates reconstruction of the LLP decay point from the crossing of track helices and the dijet direction. This reconstruction strategy assumes that the LLP decays predominantly into two jets. In contrast, the doubly charged scalar in our scenario decays into a pair of $W$ bosons, which subsequently decay inclusively. Consequently, the dijet direction does not coincide with the parent LLP direction in general. Hence the distribution of the cluster RMS variable could be very different for the situation where LLP decays to dijet and LLP decays to $W$-bosons which further decays hadronically. Furthermore, ref.~\cite{CMS_disp_jet_2019} incorporates these observables into a likelihood discriminant, while ref.~\cite{CMS_disp_jet_2020} employs them as inputs to a boosted decision tree (BDT). In both analyses, the final selection relies on stringent requirements on the likelihood or BDT score, making these multivariate discriminants among the most powerful components of the event selection. Since their performance depends sensitively on the signal topology, the published limits cannot be reliably reinterpreted for our scenario without a dedicated recasting of the analyses.

In addition to the hadronic final state, the LHC has also performed searches for LLPs with shorter decay lengths via displaced leptons, as reported in refs.~\cite{CMS_disp_lep, ATLAS_disp_lep_2024, ATLAS_disp_lep_2026}. The searches presented in refs.~\cite{CMS_disp_lep, ATLAS_disp_lep_2024} report exclusion limits in the plane of the LLP mass and proper decay length. Consequently, the quoted limits are inherently model-dependent, as they rely on the assumed production mechanism, decay topology, and signal acceptance. For instance, the benchmark signal considered in ref.~\cite{ATLAS_disp_lep_2024} corresponds to a production cross section of approximately $190~\mathrm{fb}$ for an LLP mass of $300~\mathrm{GeV}$ after including the leptonic branching fractions of the $W$ bosons, whereas the corresponding production cross section in our scenario is only about $12~\mathrm{fb}$ even before accounting for the leptonic decays of the $W$ bosons. Furthermore, the search described in ref.~\cite{CMS_disp_lep} targets LLPs decaying exclusively into displaced dilepton pairs, which differs significantly from the decay topology of the doubly charged scalar considered in this work. Therefore, a meaningful reinterpretation of these searches would require a dedicated detector-level simulation and recasting within the framework of our model, which lies beyond the scope of the present study. Although ref.~\cite{ATLAS_disp_lep_2026} presents its results as upper limits on the LLP production cross section, the analysis assumes a production mechanism and event topology that differ substantially from those relevant for our scenario. Owing to the suppressed leptonic branching fractions of the $W$ boson and the comparatively small production cross section arising from the purely electroweak production mechanism of the doubly charged scalar, as discussed above, displaced-lepton searches are expected to have substantially reduced sensitivity to our signal. Consequently, these searches are likely to impose only weak constraints, if any, on the parameter space considered in this work.

\begin{figure}[htbp!]
         \centering
         \includegraphics[width=\textwidth,]{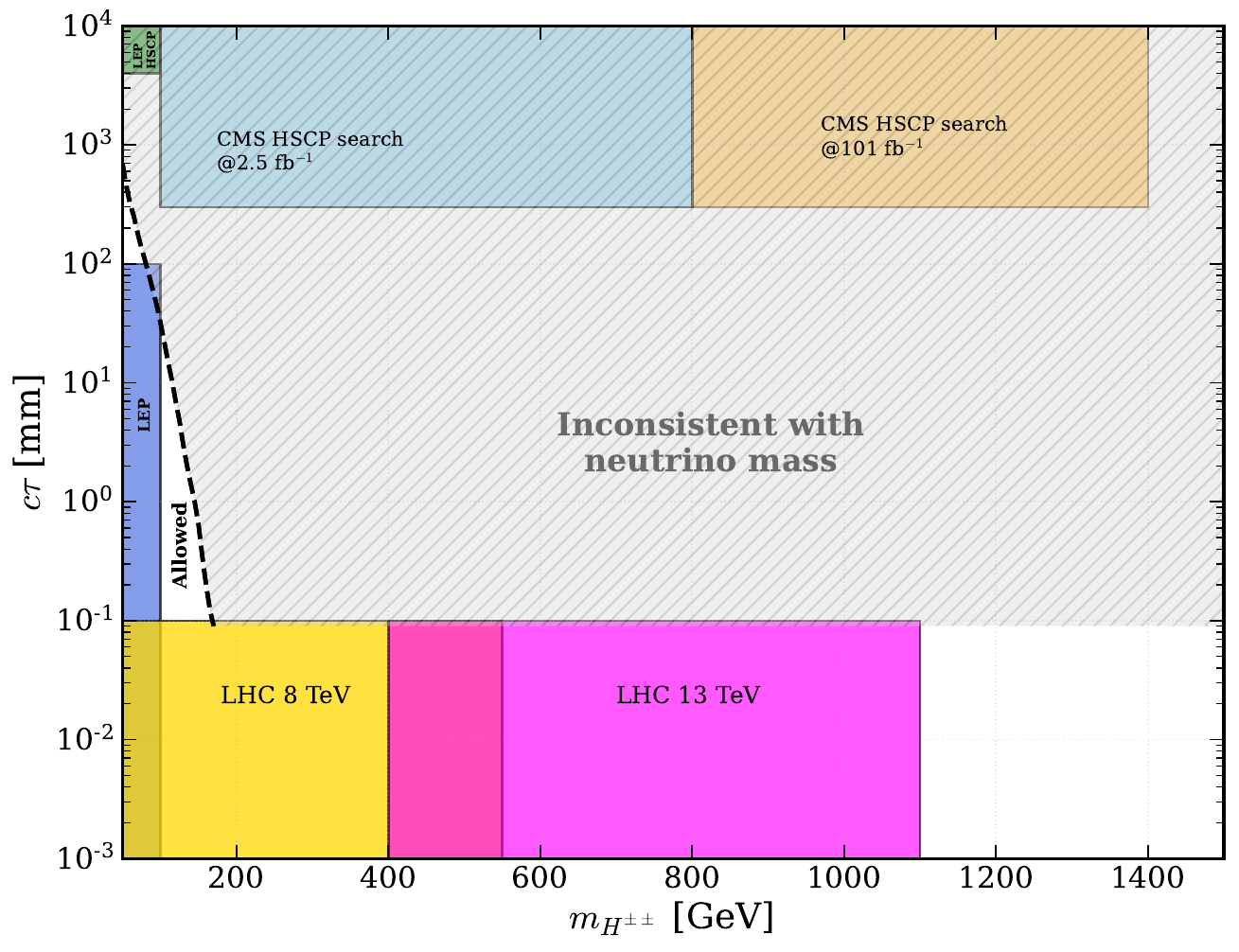}
    
         \caption{Region of parameter space in the $m_{H^{\pm\pm}} - c\tau$ plane that has been explored in experiments to date when a doubly charged scalar couples predominantly to leptons. The blue region is excluded by~\cite{OPAL_LEP}, the yellow region is excluded via~\cite{ATLAS_2014}, the magenta region is excluded via~\cite{ATLAS_2022}, the green region is 
         excluded by~\cite{DELPHI_HSCP}, the sky-blue region is excluded 
         by~\cite{cms_2016} and the orange region is excluded 
         by~\cite{cms_dedx_2025}. The dotted line represents the upper limit of $c\tau$ for different values of $m_{H^{\pm\pm}}$ }
         \label{sum_yuk}
\end{figure}%

\begin{figure}[htpb]
         \centering
    \includegraphics[width=0.95\textwidth,]{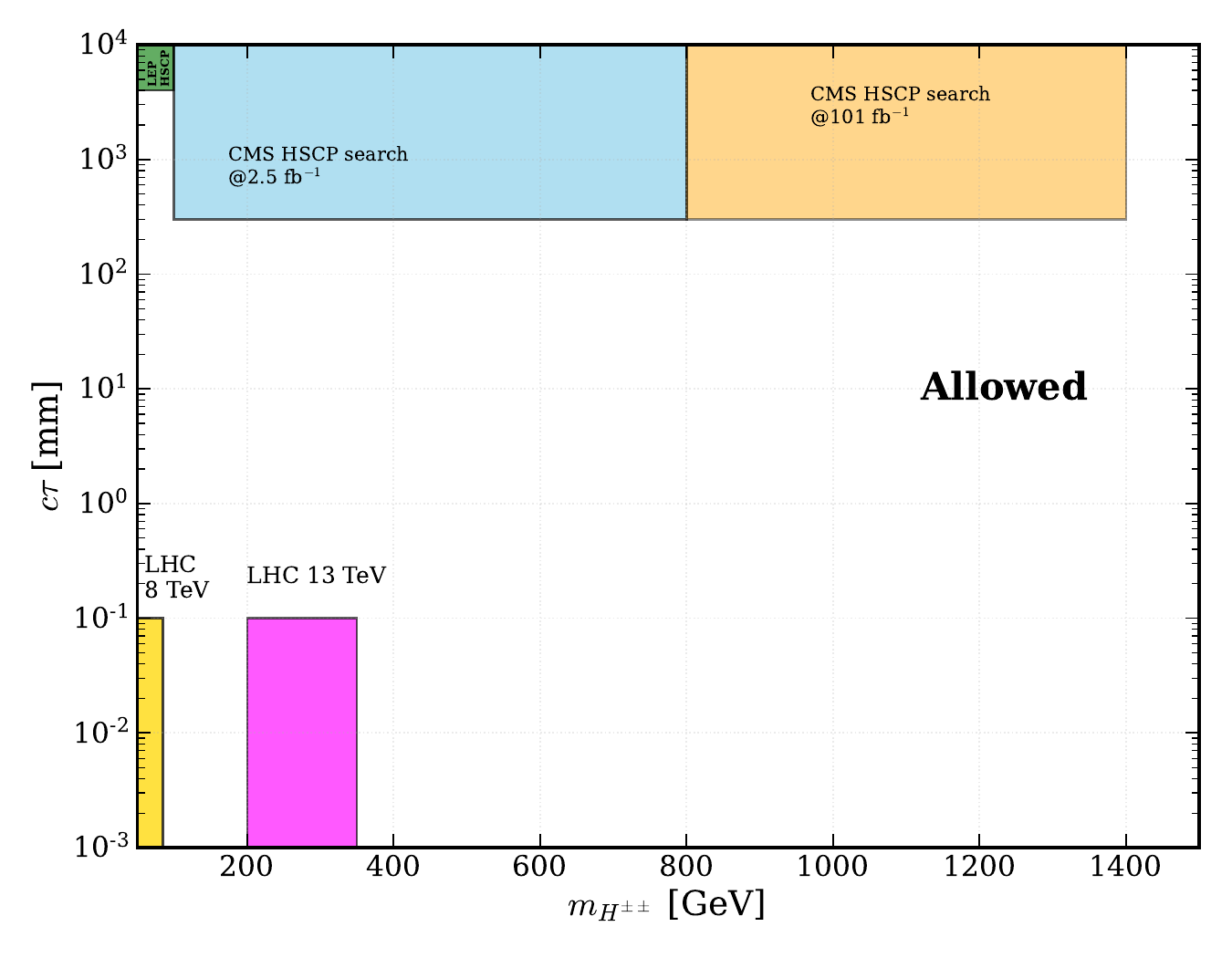}
         \caption{Region of parameter space in the $m_{H^{\pm\pm}} - c\tau$ plane that has been explored in experiments to date when a doubly charged scalar couples predominantly to gauge bosons. The yellow region is excluded via a recast of the anomalous like-sign dilepton search by ATLAS in 
         Run-1~\cite{ATLAS_2014,recast}, the magenta region is excluded via~\cite{atlas_2021}, the green region is 
         excluded by~\cite{DELPHI_HSCP}, the sky-blue region is 
         excluded by~\cite{cms_2016} and the orange region 
         is excluded by~\cite{cms_dedx_2025}}
         \label{sum_noyuk}
\end{figure}%

Here we summarize the constraints imposed by different experimental searches on our scenario.  Though most of the searches for heavy stable charged particles at ATLAS depend on the signal model and signal topology considered because the trigger is based on missing energy, the search for multi-charged particles\cite{atlas_mcp} excludes a doubly charged scalar with mass up to $1~\mathrm{TeV}$. But since ref-\cite{atlas_mcp} does not explicitly mention the range of proper decay length it is sensitive to, we are not displaying its bound on $m_{H^{\pm\pm}} - c\tau$ plane. It is worth restating that, this search requires the LLP to reach the muon chamber, hence expected to be sensitive to higher decay length. In contrast, the HSCP searches by CMS are applicable to a broader class of signal models, and also constrain the DY pair-production cross section for the LLPs. Besides HSCP searches, CMS has also performed searches for displaced vertices, but these searches are signal-topology-dependent and thus require detailed recasting to derive constraints on different models. Derivations of these constraints are neither of interest nor within the scope of the present work. Fig.~\ref{sum_yuk} and fig.~\ref{sum_noyuk} summarize the regions in the $m_{H^{\pm\pm}}$-$c\tau$ plane that have been probed experimentally for two representative scenarios: (i)~when the doubly charged scalar couples to both leptons and gauge bosons, 
and (ii)~when it couples exclusively to gauge bosons.  Here we have marked the regions that have been ruled out experimentally. In both figures, we have considered the prompt limit to be imposed below $0.1~\mathrm{mm}$ and, as it has been shown in fig.~\ref{prmpt_eff}, a $c\tau$ of $0.1~\mathrm{mm}$ corresponds to 60\% efficiency. Since the PV reconstruction efficiency is usually kept higher than this, our consideration remains conservative. In fig.~\ref{sum_yuk}, although a region of parameter space remains experimentally unconstrained, we do not find any parameter point with a delayed decay of the doubly charged scalar for masses above $150~\mathrm{GeV}$ within this otherwise allowed region. Note that both of these figures are representative of the mass exclusions reported by different experimental collaborations so far.

\section{Collider analysis}
\label{col}

As mentioned in section~\ref{constraint}, the search for a doubly charged scalar is efficient in two regimes of decay lengths, i.e., $c\tau > \mathcal{O}(100~\mathrm{mm})$ and $c\tau < \mathcal{O}(0.1~\mathrm{mm})$ via HSCP searches and prompt searches,  respectively. 

\begin{figure}[htb!]
\includegraphics[width=0.9\textwidth]{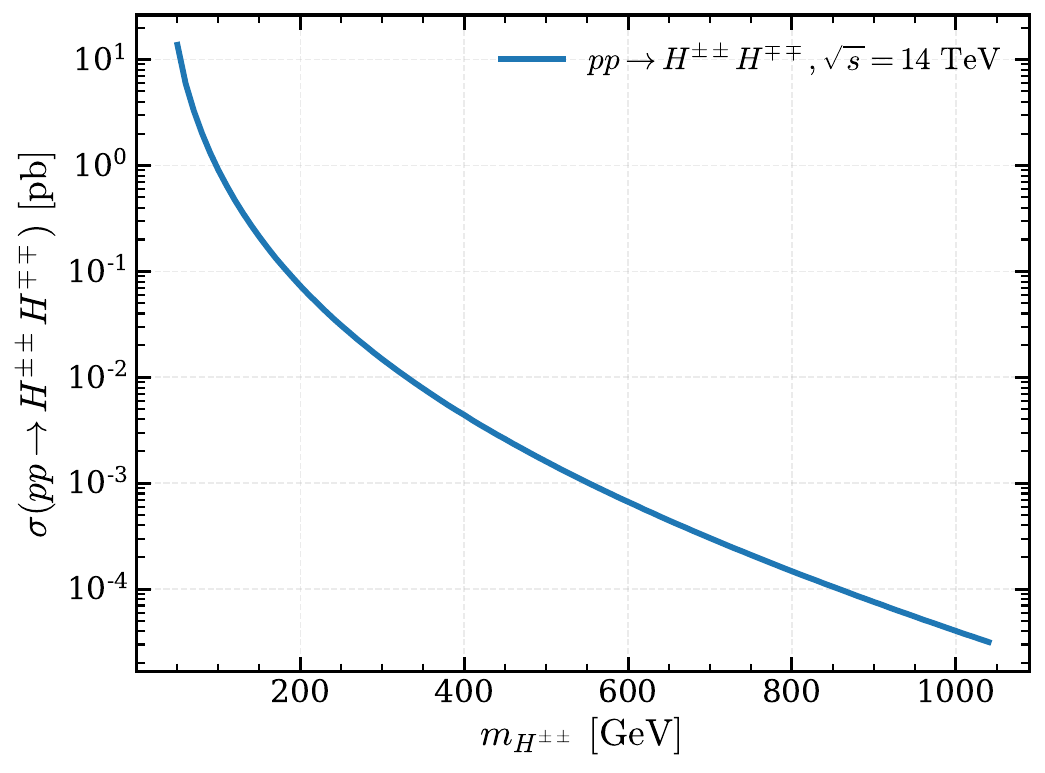}
         \caption{Variation of the production cross section of the doubly charged scalar via the DY process at $\sqrt{s} = 14$~TeV. }
         \label{dyd}
\end{figure}

Here we will focus on the parameter space where the decay length of the doubly charged scalar is in the range $\mathcal{O}(10~\mathrm{mm})~\text{to}~\mathcal{O}(100~\mathrm{mm})$. As already depicted in fig.~\ref{llp_const}, both prompt and HSCP searches lose sensitivity in this range of decay length. Moreover, in this regime the searches become signal topology dependent and hence ask for a dedicated strategy. Here we shall discuss a displaced vertex search strategy for such decay lengths, taking into account both the situations where the doubly charged scalar decays only into same-sign $W$ bosons and the doubly charged scalar decays into same-sign dileptons as well as same-sign $W$ bosons. The production mechanism for the signal has been considered to be the DY pair production of the doubly charged scalar. The consideration of DY pair-production as the production mechanism for the signal frees the production cross section from significant model dependence.  The signal cross section 
of the DY pair-production at $14~\mathrm{TeV}$ $pp$-collision is 
shown in fig.~\ref{dyd}.

\subsection{Simulation of signal and background}
\label{simulation}

As shown in fig.~\ref{regplot}, in the presence of the Yukawa coupling of the triplet scalar, only in the low-mass region, i.e., 
$100~\mathrm{GeV}-150~\mathrm{GeV}$, the doubly charged scalar behaves as an LLP, while in the absence of the Yukawa coupling, in a larger portion of the parameter space the doubly charged scalar appears to be an LLP. We shall consider a benchmark in the low-mass region and three benchmarks in the high-mass region, i.e., above the $WW$ threshold. 

The signals have been generated at leading order with the parton distribution function \textsc{NNPDF23LO}~\cite{nnpdf}. With the Feynrules model 
file~\cite{decoup,gmfr} the pair-production of the charged scalars has been performed using \textsc{MadGraph5\_aMC@NLO}~\cite{mg5} , and the scalars have been decayed in \textsc{Pythia8}~\cite{pythia}.
For the cases where charged scalars decay into vector bosons, the further decay of the vector bosons has been taken to be inclusive. As a consequence, the final state of the signal can be hadronic, semi-leptonic, and leptonic. For the signal as well as all the backgrounds, showering and hadronization have been done with \textsc{Pythia8} and the detector response has been simulated by \textsc{Delphes-3.5.0}~\cite{delphes} using the default CMS Delphes card for the HL-LHC.

Since the final states of our interest feature hadronic decay of vector bosons, the QCD multijet processes constitute a significant portion of backgrounds. After hadronization, this can produce baryons or mesons with a lifetime $\mathcal{O}(1~\mathrm{ps})$, and at LHC boosts these particles can traverse a distance $\mathcal{O}(1~ \mathrm{mm})$ and hence can potentially mimic the signature of a long-lived particle. As the long-lived mesons or baryons of this kind primarily originate from the hadronization of heavy flavor quarks, in order to populate this region well enough, we have generated the light flavor quarks and heavy flavor ones separately, and the label QCD incorporates both of them. These backgrounds have been produced using \textsc{Pythia8} with 
a generation-level cut of $20~\mathrm{GeV}$ on the $p_T$ of the final state particles of the hard process. 

Apart from QCD multijet, we have also considered backgrounds coming from $t\bar{t}$ and $V + \text{jets}$, where $V$ represents $W$ or $Z$ bosons. Here, the top quark and the vector bosons have been considered to decay inclusively. These processes have been generated with a minimum cut of $20~\mathrm{GeV}$ on the $p_T$ of the final state particles in \textsc{MadGraph5\_aMC@NLO} . 

We have generated 200M events for QCD multijet and 1M events for each of the $t\bar{t}$, $W+\text{jets}$ and $Z+\text{jets}$ samples. To use the storage more efficiently, we store only those events where there is at least one displaced vertex with invariant mass $(m_{DV})$ greater than or equal to $1~\mathrm{GeV}$. The construction of $m_{DV}$ has been discussed below.

Another source of background is instrumental in nature, and simulating it is out of the scope of this work. Future \textsc{Geant4}~\cite{geant4} simulations will be useful to accurately model such backgrounds arising from high-pileup environments ($\langle\mu\rangle \approx 140-200$). To make sure that the displaced vertex is well away from the regions of the tracker layers where material density is high, and SM particles can interact with the detector material producing displaced vertices, a material-map veto has been applied to the reconstructed displaced vertex. The material-map is based on Run-1 and Run-2 data collected at the CMS detector and has been taken from~\cite{cms_matmap}.

\textbf{Reconstruction of displaced vertex at generator level:} The presence of a displaced vertex in an event has been searched for at the generation level in \textsc{Pythia8}~\cite{pythia} after showering and hadronization. For each event, the final state particles with $L_{xy} > 0.5~
\text{mm}$ have been collected, and their production vertices are stored. Further, from this list of particles, only the charged particles have been retained, as only these particles can be seen at the tracker. For each charged particle with $p_T > 1~ \text{GeV}$, $|\eta| < 2.5$ and $|d_0| > 0.3~\text{mm}$, the vertex list has been looped over, and for the vertex where the distance between the potential track and vertex is minimum, the track has been associated with the vertex. If two such vertices have distance $\le 0.01~\text{mm}$, they have been merged. Finally, the invariant mass of the thus reconstructed displaced vertex, denoted as $m_{DV}$, has been calculated using the momentum of the associated charged particles, and only those events have been stored that have at least one displaced vertex with $m_{DV} > 1~ \text{GeV}$. This criterion has been mentioned as `Gen. selection' in Table~\ref{uptopre}.

\textbf{Reconstruction of displaced vertex after detector simulation:} \label{mdv_gen} After detector simulation with \textsc{Delphes-3.5.0}~\cite{delphes}, events were initially selected via a pre-selection criterion. The pre-selection criteria demand an event to satisfy any one of the following two requirements: (1)~presence of at least two jets with $p_T > 40~\mathrm{GeV}$ and $|\eta|< 2$, or (2)~a displaced lepton with $|d_0| > 0.5~\mathrm{mm}$, $p_T > 20~\mathrm{GeV}$ and $|\eta| < 2.5$. Then the events were further divided into three categories, namely hadronic, semi-leptonic, and leptonic, to efficiently capture the signal for different decay modes of the doubly charged scalar.

\begin{itemize}
    \item In the hadronic category two more jets with $p_T > 20~\mathrm{GeV}$ and $|\eta|< 2$ were demanded. All possible combinations of 4 jets were constructed out of the selected jets. Then tracks were associated with every four-jet combination. The tracks considered here have transverse momentum $> 1~\mathrm{GeV}$ and $|\eta| < 2.5$. For a given combination of 4 jets, a track was associated with a jet if the $\Delta R$ between the track and jet is minimum and $< 0.5$. 

    After associating tracks to each jet, a collection of displaced tracks was made out of the tracks having $100~ \mathrm{mm} > |d_0| > 0.5~\mathrm{mm}$. These displaced tracks were used to reconstruct the displaced vertex.

    The origin of these displaced tracks was assumed to be the candidates for the displaced vertex, and if any two displaced vertex candidates are closer than $1~\mathrm{mm}$, they are merged. For the finally surviving displaced vertex candidates, tracks were associated with the vertices if the distance between the track and the displaced vertex is minimum and lower than $1~\mathrm{mm}$.

    For further analysis, the displaced vertices with track multiplicity $> 2$ have been used.

    \item Events having two jets with the above-mentioned $p_T$ and $\eta$ and exactly two opposite-sign displaced leptons have been selected in the 
    semi-leptonic category. 
    For a selected event, all possible dijet candidates have been formed, and tracks were associated with each candidate following the same procedure as explained above.
    The displaced tracks associated with the jets and the leptons have been used to reconstruct the displaced vertex.

    \item In the leptonic category, two same-sign leptons are required with $|d_0| > 0.5~\mathrm{mm}$, $p_T > 20~\mathrm{GeV}$ and $|\eta| < 2.5$. Only those events have been retained where both of these leptons have been found to originate from the same vertex. This has been ensured by demanding that the starting point of one lepton be within $1~\mathrm{mm}$ of the starting point of another.
\end{itemize}

\subsection{Cut-based analysis}

As has already been emphasized, the scenario where the complex triplet couples to leptons is of relevance here only in the low-mass region, i.e., below $150~\mathrm{GeV}$. A huge parameter space that asks for a dedicated search strategy opens up when the doubly charged scalar exclusively decays into same-sign $W$ bosons. Hence, we will have two distinct signatures in the low-mass regime: (1)~The DY pair-production of a doubly charged scalar and its subsequent decay into 
a $W$ boson pair and (2)~The DY pair-production of a doubly charged scalar and its subsequent decay into same-sign dileptons. Only the first kind appears above the $WW$ threshold. Hence, after taking into account all possible signal topologies, the events can be classified into three categories: fully hadronic final state, semi-leptonic final state, and fully leptonic final state. In order to maximize the efficiency of the signal selection, the analyses have been done in three categories geared towards selecting each topology. Hence, these three analyses attempt to reconstruct three distinct types of vertices as discussed below.

\begin{enumerate}
\item \textbf{Type-1:} The doubly charged scalar decays into two same-sign leptons, via $H^{\pm\pm} \to \ell^{\pm} \ell^{\pm}$ or 
$H^{\pm\pm} \rightarrow W^\pm W^\pm \rightarrow \ell^{\pm} \nu_{\ell} \, \ell^{\prime \pm}\nu_{\ell^{\prime}}$. The first kind appears in the $100-150~\mathrm{GeV}$ range only, while the second one appears for the whole mass range. The leptonic analysis aims to reconstruct this vertex.

\item \textbf{Type-2:} The doubly charged scalar decays into two same-sign $W$ bosons, and both of them decay hadronically: $H^{\pm\pm} \rightarrow W^\pm W^\pm \rightarrow \text{hadrons}$. The hadronic analysis reconstructs this vertex.
    
\item \textbf{Type-3:} The doubly charged scalar decays into two same-sign $W$ bosons, and one of them decays hadronically and the other leptonically: $H^{\pm\pm} \rightarrow W^\pm W^\pm \rightarrow q\bar{q}^{\prime} \, \ell^\pm \, \nu_\ell$. The semi-leptonic analysis targets these types of vertices.
\end{enumerate}

\begin{figure}[htbp!]
	\centering
    \subfloat[]{\label{ht_eff}\includegraphics[width=0.8\textwidth]{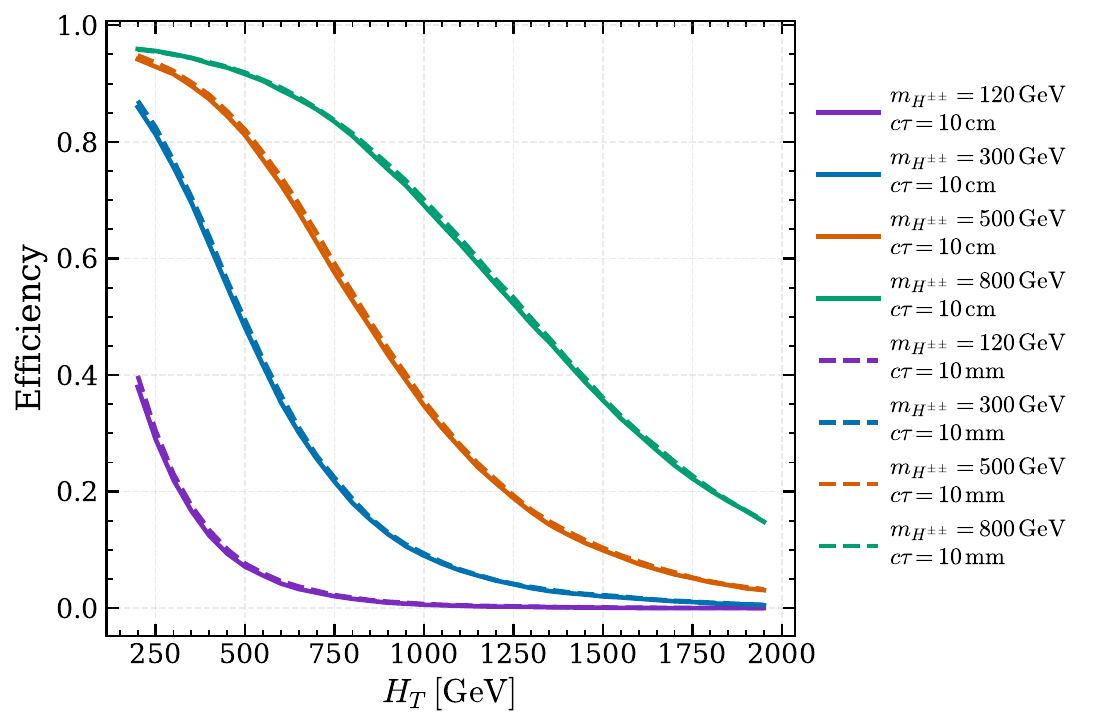}}\\
    \subfloat[]{\label{disp_lep_eff}\includegraphics[width=0.8\textwidth]{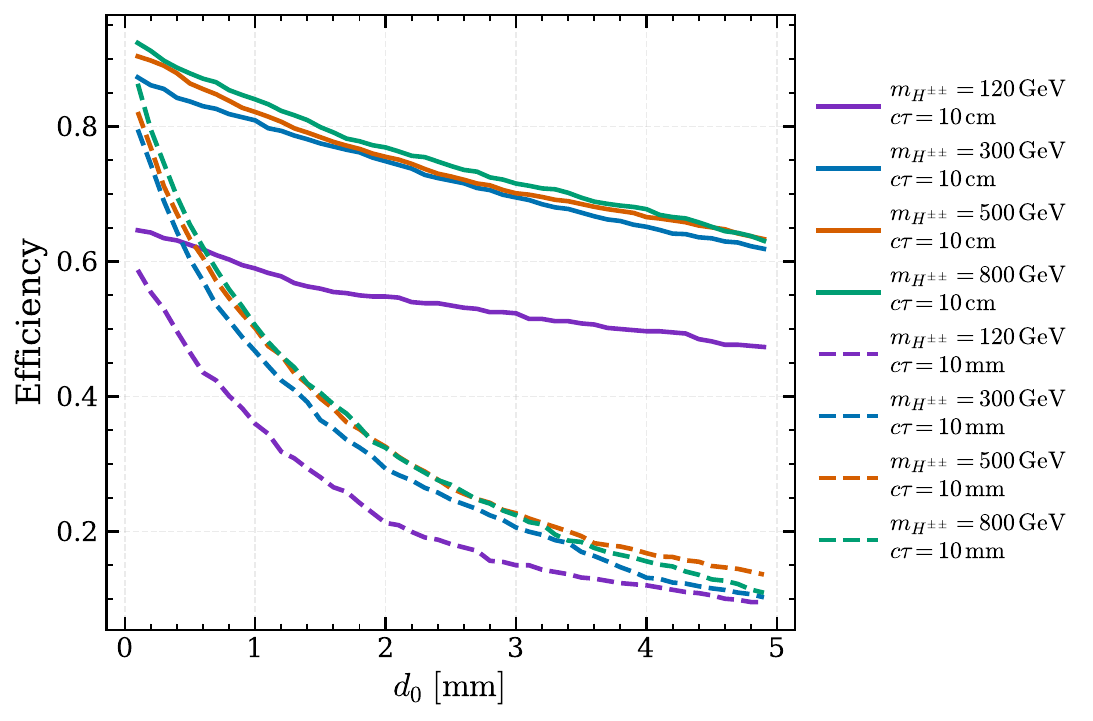}}
    \caption{Top: Efficiency of $H_T$ cut for different mass points. Bottom: Efficiency of displaced lepton selection, which can be used at the trigger level, for different masses and decay lengths. The x-axis notes the absolute values of $d_0$. } 
	\label{trig_eff}
\end{figure}

As is quite obvious from the nomenclature, the hadronic analysis is geared towards capturing the vertices of type-2, the semi-leptonic analysis takes into account the vertices of type-3, and the leptonic analysis reconstructs the vertices of type-1. Categorizing the vertices in these three classes ensures a better reconstruction of the displaced vertex for such a signal where the LLP has multiple decay modes and selecting events via a single trigger may underestimate the signal efficiency. 

For a search for this type of signal, the experimental collaborations should introduce dedicated L1T and HLT based on displaced jets and displaced leptons, for the best possible signal efficiency. Some of the current LLP searches use $H_T$ triggers, so in Fig.~\ref{ht_eff} we have shown the variation of our signal efficiency with $H_T$ for different masses of the decaying scalar. For this plot, we have required an event to have at least two jets with $p_T > 40~\mathrm{GeV}$ and $|\eta| < 2$, and then the $H_T$ has been calculated as the scalar sum of the $p_T$ of all jets that satisfy these $p_T$ and $\eta$ criteria. As evident from fig.~\ref{ht_eff}, this trigger is not very efficient for lower mass points. So, we do not recommend using the usual $H_T$ trigger for such searches. We expect that at the HL-LHC, it will be possible to trigger on our signal events by using displaced-track trigger algorithms. This will ensure high online signal acceptance while maintaining manageable trigger rates under high-pileup conditions. 

In this work, we assume a displaced-object trigger operating at high eﬀiciency for events passing the selection shown in fig.~\ref{cutflow}. As shown there, we have selected events based on the $p_T$ and $\eta$ of jets, and also $p_T$, $\eta$, and displacement of leptons. Also note that, since we are categorizing events into three non-overlapping categories, our analysis will benefit from multiple triggers, each targeted towards a specific final state. At the very least, one hadronic trigger and one leptonic trigger will be needed. 
Fig.~\ref{disp_lep_eff} shows the efficiency of the following cuts intended to be used at the trigger level: at least two leptons with $p_T > 20~\mathrm{GeV}, \, |\eta| < 2.5~ \text{and}~ 100~\mathrm{mm} > |d_0| > 0.5~\mathrm{mm}$ for different masses and different decay lengths. The efficiency shown in the figure is relative to the number of events where at least 2 leptons were present.

Fig.~\ref{trig_eff} has been obtained for the scenario where the doubly charged scalar decays exclusively into $W$ bosons. In the scenario where $H^{\pm\pm}$ couples to leptons as well as $W$ bosons, the efficiency of the $ H_T$-based trigger is not expected to change, as this trigger is dependent on the $p_T$ of the jets. But for the displaced lepton trigger, the efficiency is expected to increase with respect to what is shown in fig.~\ref{disp_lep_eff} for $120~\mathrm{GeV}$. 

Events selected by the hadronic trigger and not the leptonic one are passed to the hadronic analysis; events selected by both triggers and the leptons are of opposite charges, and the events undergo the semi-leptonic analysis. Finally, if an event contains at least two same-sign leptons, the corresponding events go through the leptonic analysis. A flow-chart summarizing the cuts implemented in each analysis is given in fig-\ref{cutflow}. Note that events with one jet and one lepton though passes pre-selection, does not enter any of the final analysis mode as it does not correspond to any possible final state topology for the doubly charged scalar.

\vspace{0.5cm}

\begin{figure}
\centering
    
\begin{tikzpicture}[
    node distance=10mm and 0.02mm,
    >=Latex,
    box/.style={
        draw,
        rounded corners,
        align=center,
        minimum width=5mm,
        minimum height=2mm,
        font=\small
    }
]
\node[box, minimum width=8cm] (top) {\textit{Pre-selection}:
Events with at least two jets with \\[0.5mm]
$p_{T}\ge 40~\mathrm{GeV}$ and $|\eta|\le 2$ \\ or \\ at least one lepton with $p_T > 20~\mathrm{GeV}$, \\ $|\eta|<2.5$, $|d_0| > 0.5~\mathrm{mm}$ and $|d_0| < 100 ~\mathrm{mm}$
};
\node[box, below left=of top] (left) {
Events with at least 2 jets \\
and at most 1 lepton};

\node[box, below=of left] (3rd left) {\textbf{Hadronic analysis}\\
\textit{Cut-1}: Two more jets \\
with $p_T \ge 20~\mathrm{GeV}$. \\
\textit{Cut-2}: At least one \\ displaced vertex \\ reconstructed
from tracks \\ with $0.5~\mathrm{mm} \le d_0 \le 100~\mathrm{mm}$.\\
\textit{Cut-3}: The displaced \\ vertex must  have \\
$> 2$ displaced tracks.\\
\textit{Cut-4}: Vertex must survive \\ material-map veto.\\
\textit{Cut-5}: $m_{DV} \ge 20~ \mathrm{GeV}.$
};

\node[box, below=of top] (middle) {
Events with $>= 2$ jets \\
and exactly 2 leptons\\
(must have opposite signs)};

\node[box, below=of middle] (3rd middle) {\textbf{Semi-leptonic analysis}\\
\textit{Cut-1}: Exactly two opposite sign \\ 
leptons with $p_T > 20~ \mathrm{GeV}$, ${|\eta|}<2.5$ \\
 and $0.5~\mathrm{mm}< |d_0| < 100~\mathrm{mm}$.\\
\textit{Cut-2}: At least one displaced vertex \\ reconstructed
from tracks \\ with
$0.5~\mathrm{mm} \le |d_0| \le 100~\mathrm{mm}$.\\
\textit{Cut-3}: The displaced vertex \\ must have
$> 2$ tracks.\\
\textit{Cut-4}: Vertex must survive \\
material-map veto.\\
\textit{Cut-5}: $m_{DV} \ge 20~ \mathrm{GeV}.$
};

\node[box, below right=of top] (right) {
Events with $\ge 2$\\
leptons\\
(at least 2 same-sign)};

\node[box, below=of right] (3rd right) {\textbf{Leptonic analysis}\\
\textit{Cut-1}: At least two \\ same-sign leptons \\
with $p_T > 20~ \mathrm{GeV}$, ${|\eta|}<2.5$ \\
and $0.5~\mathrm{mm}< |d_0| < 100~\mathrm{mm}$.\\
\textit{Cut-2}: Both leptons must come \\
from the same displaced vertex.\\
\textit{Cut-3}: The displaced vertex \\ must have $>= 2$ tracks.\\
\textit{Cut-4}: Vertex must survive \\
material-map veto.\\
\textit{Cut-5}: $m_{DV} \ge 20~ \mathrm{GeV}.$
};
\draw[->, thick] (top.south) -- (left.north);
\draw[->, thick] (top.south) -- (middle.north);
\draw[->, thick] (top.south) -- (right.north);
\draw[->, thick] (right) -- (3rd right);
\draw[->, thick] (left) -- (3rd left);
\draw[->, thick] (middle) -- (3rd middle);
\label{cutflow}
\end{tikzpicture}
\caption{Flow-charts of the cuts implemented in this analysis. The top box defines the pre-selection criterion. Middle row boxes: Criteria that classify pre-selected events for 3 analyses. Bottom boxes: cut-flow implemented in hadronic, semi-leptonic, and leptonic analyses, respectively, from left to right.}
      \label{cutflow}
\end{figure}
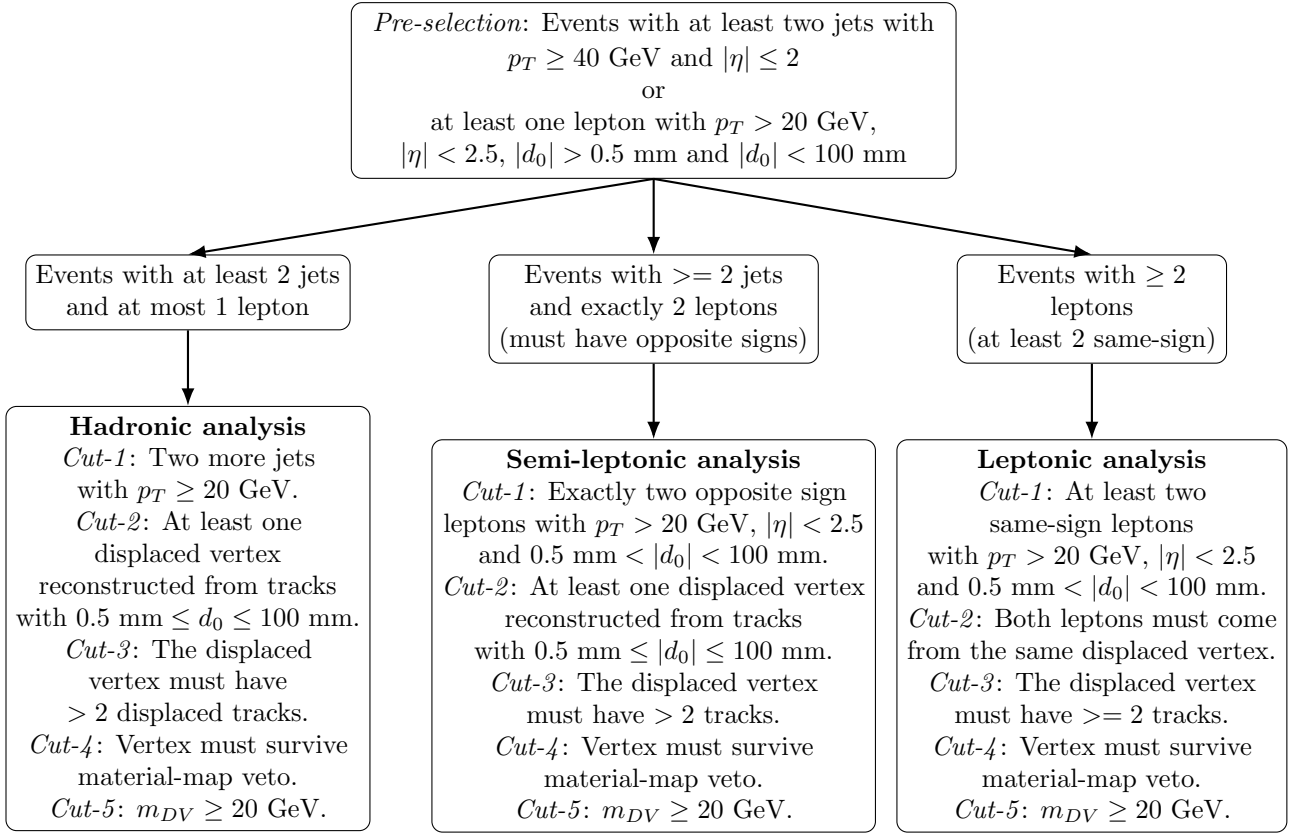

\begin{figure}[htbp!]
	\centering
    \subfloat[]{\label{300h}\includegraphics[width=0.35\textwidth]{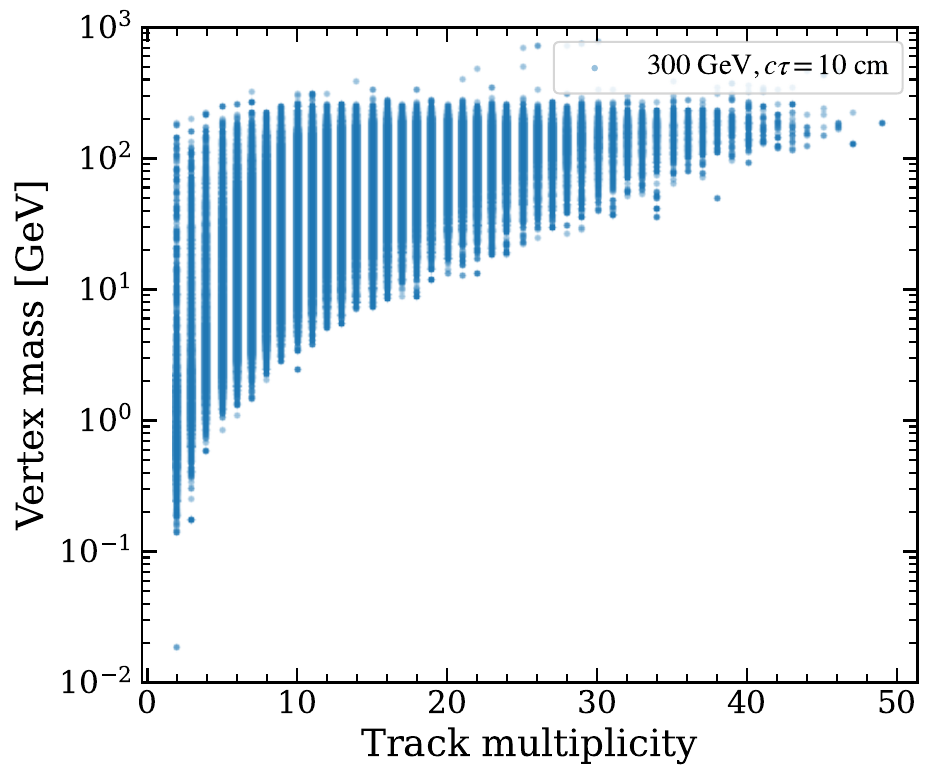}}
    \subfloat[]{\label{300sl}\includegraphics[width=0.35\textwidth]{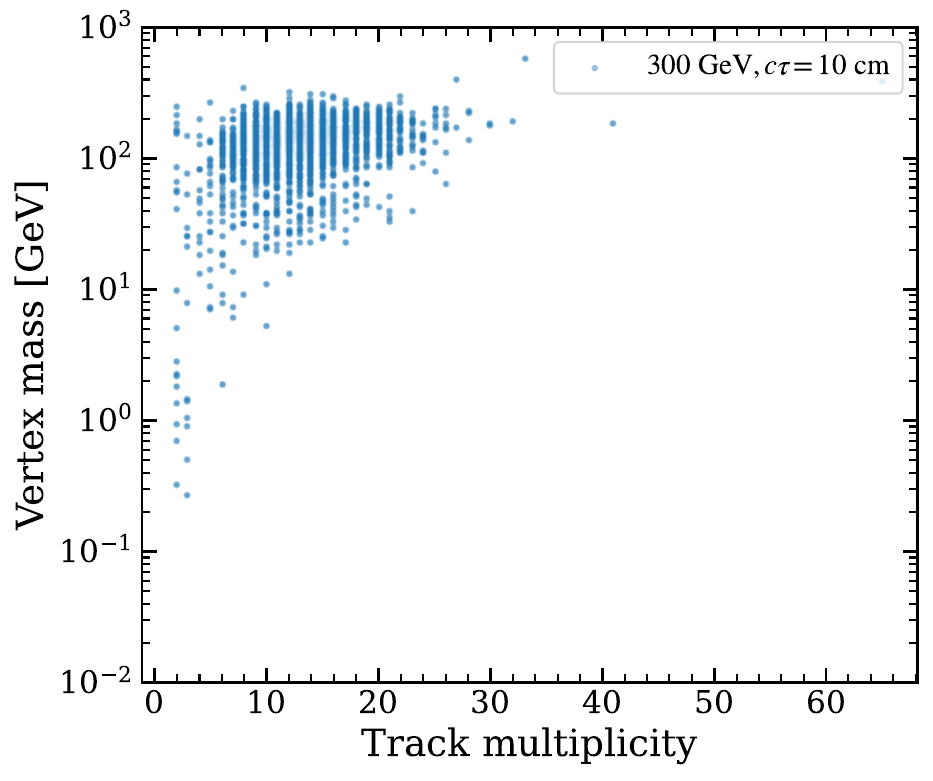}}
	\subfloat[]{\label{300l}\includegraphics[width=0.35\textwidth]{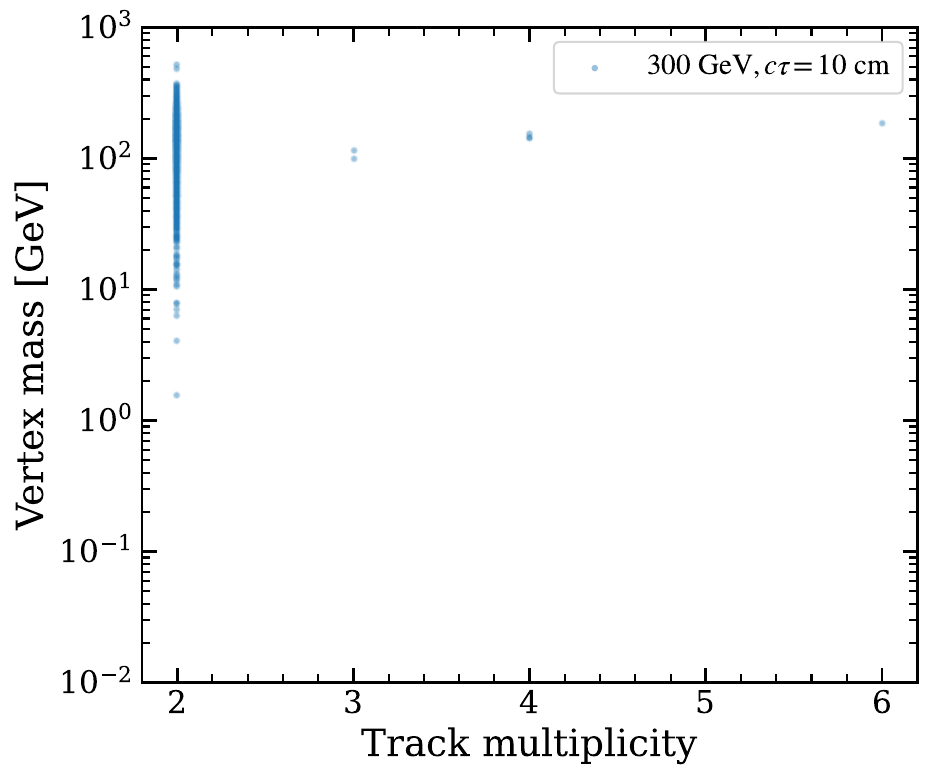}}\\
    \subfloat[]{\label{800h}\includegraphics[width=0.35\textwidth]{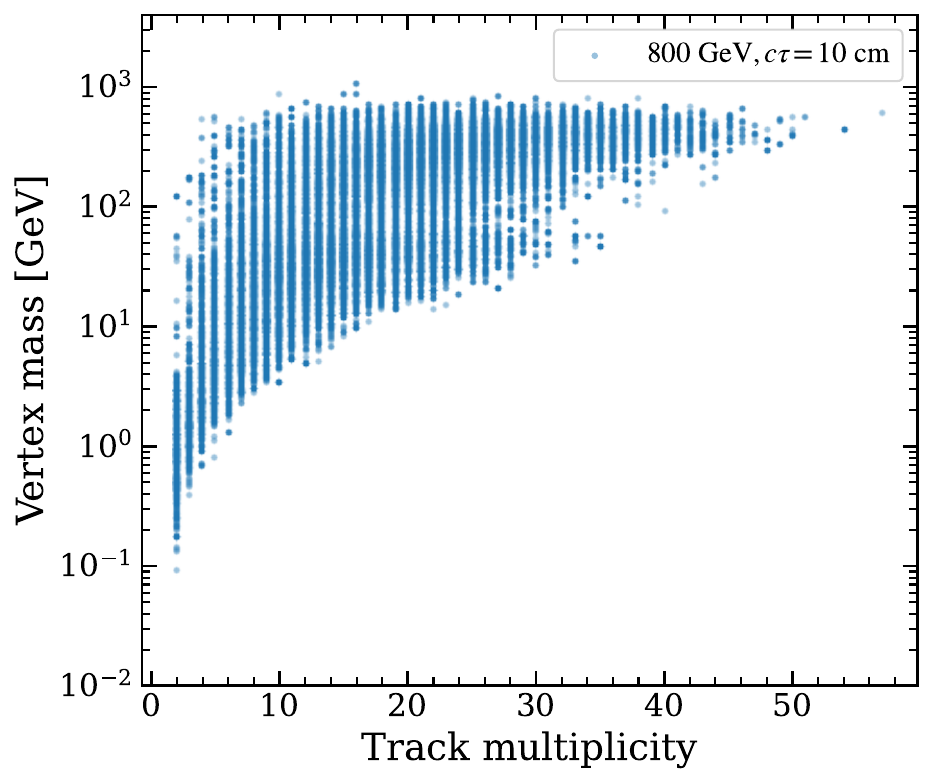}}
    \subfloat[]{\label{800sl}\includegraphics[width=0.35\textwidth]{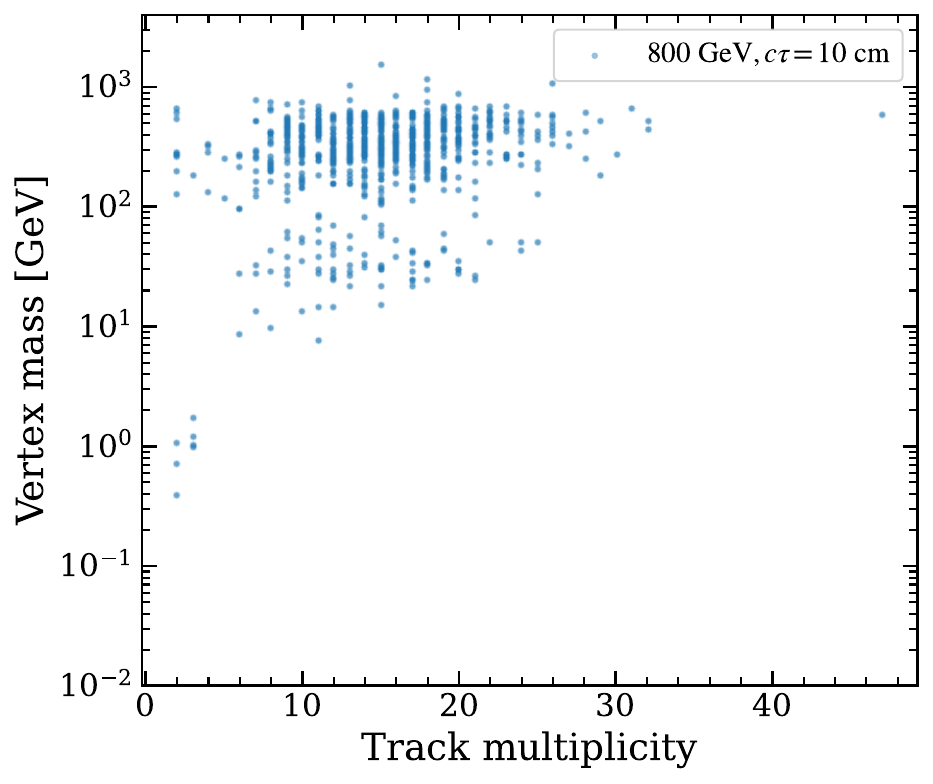}}
	\subfloat[]{\label{800l}\includegraphics[width=0.35\textwidth]{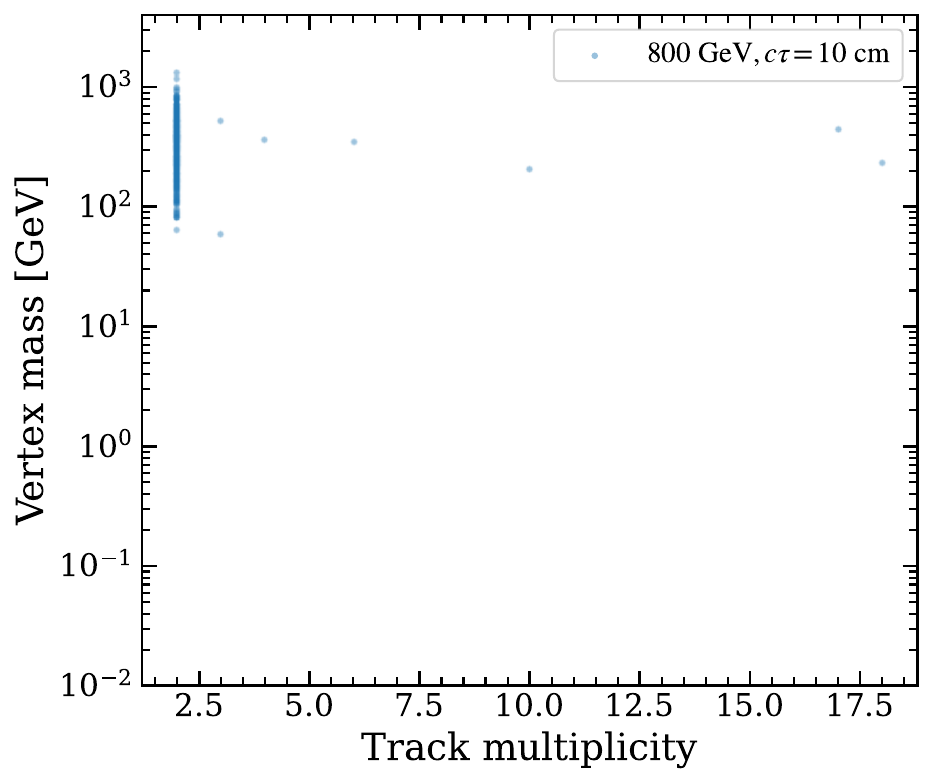}}
	\caption{Track multiplicity vs invariant mass reconstructed from the tracks associated with the displaced vertex $(m_{DV})$ for a scalar mass of 800~GeV~(bottom) and 300~GeV~(top). Left panel: Hadronic channel, middle panel: semi-leptonic channel, right panel: leptonic channel.} 
	\label{trk_mult_vm_sig}
\end{figure}

\begin{figure}[htpb!]
	\centering
    \subfloat[]{\label{vmjj}\includegraphics[width=0.35\textwidth]{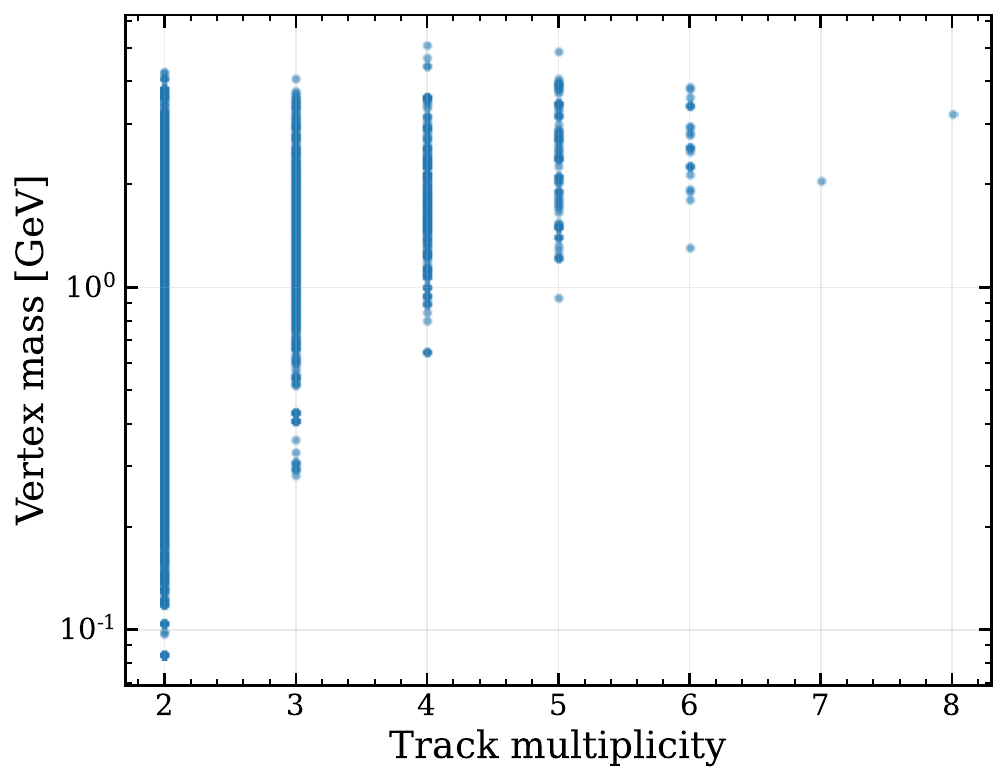}}
    \subfloat[]{\label{vmtt}\includegraphics[width=0.35\textwidth]{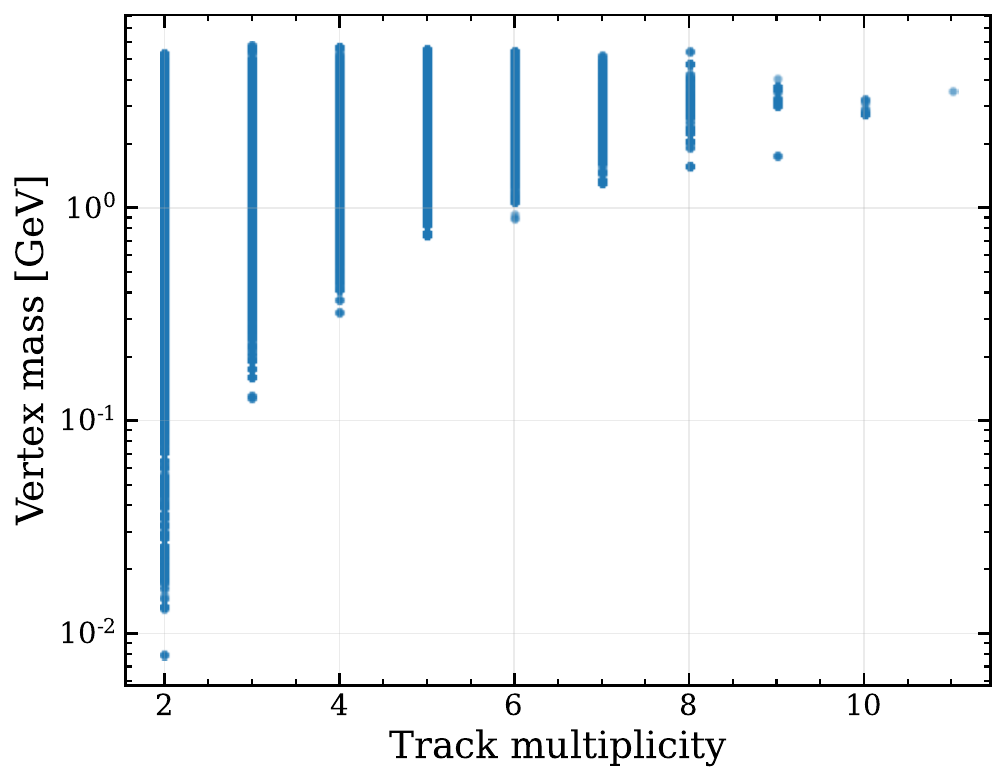}}
	\subfloat[]{\label{vmvj}\includegraphics[width=0.35\textwidth]{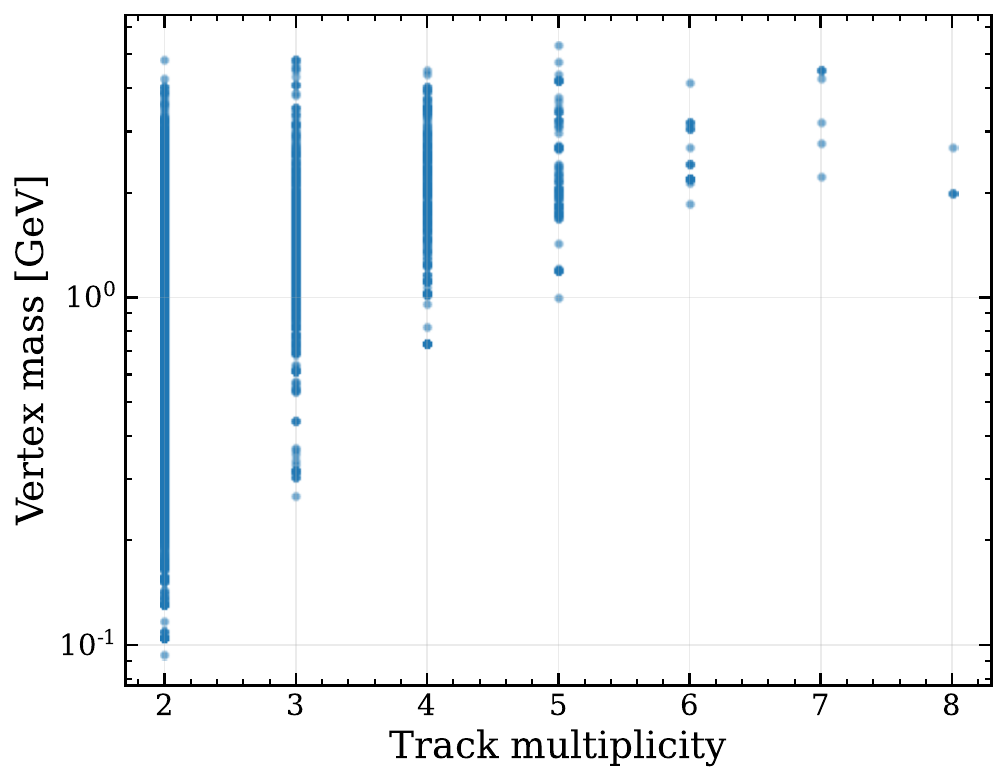}}\\
    
	\caption{ Track multiplicity vs invariant mass reconstructed from the tracks associated with the displaced vertex $(m_{DV})$ for all backgrounds, i.e., QCD,  $t\bar{t}$  and V$+$jets from left to right for hadronic analysis.} 
	\label{trk_mult_vm_bkg}
\end{figure}

Fig.~\ref{trk_mult_vm_sig} and fig.~\ref{trk_mult_vm_bkg} show track multiplicity vs invariant mass of the displaced vertex as constructed from the displaced tracks. 

As summarized in the cut-flow, the hadronic analysis attempts to reconstruct the displaced vertex associated with the decay chain $H^{\pm\pm} \rightarrow W^\pm W^\pm \rightarrow \mathrm{hadrons}$.
In this channel, both $W$ bosons decay hadronically, resulting in four jets that originate from the displaced decay vertex. The semi-leptonic analysis, on the other hand, reconstructs the displaced vertex from the decay chain
$H^{\pm\pm} \rightarrow W^\pm W^\pm \rightarrow q\bar{q}^{\prime} \, \ell^\pm \, \nu_\ell$, 
where one $W$ boson decays hadronically and the other decays leptonically. Consequently, this analysis considers one jet and one charged lepton as the visible decay products associated with the displaced vertex. Finally, the leptonic analysis targets the decay chains
$H^{\pm\pm} \rightarrow W^\pm W^\pm
\rightarrow \ell^\pm\nu_\ell,\ell^{\prime\pm}\nu_{\ell'}$, and $H^{\pm\pm} \rightarrow \ell^\pm \ell^\pm$. In this case, the displaced vertex is reconstructed solely from the two displaced charged leptons, while the neutrinos remain undetected. The three analyses therefore exhibit a clear hierarchy in their expected track multiplicity. The hadronic channel, with four jets originating from the displaced decay, has the highest track multiplicity. In the semi-leptonic channel, only one $W$ boson decays hadronically, resulting in fewer charged-particle tracks than in the fully hadronic case. The leptonic channel has the lowest track multiplicity, as the displaced vertex is reconstructed from only two charged leptons. Thus, moving from the hadronic to the semi-leptonic and finally to the leptonic analysis, the track multiplicity progressively decreases, reaching a minimum of two reconstructed tracks in the leptonic channel, as shown in fig.~\ref{trk_mult_vm_sig}. While there are enough signal events left for each category, requiring a displaced lepton can significantly reduce the background. There was no event surviving the requirement of a displaced lepton in the limited MC sample that we could generate. Hence, in 
fig.~\ref{trk_mult_vm_bkg} we only show track multiplicity vs invariant mass of the displaced vertex for hadronic analysis for the backgrounds. For backgrounds, the invariant mass reconstructed from the displaced tracks lies well below $10~\mathrm{GeV}$, while for the signal it could be hundreds of GeV, making $m_{DV}$ a strong discriminator between signal and background. In the background, the displaced vertex can arise from the hadronization of quarks into relatively long-lived mesons (such as $K, B, D$) or baryons (like $\Lambda~\text{or}~ \Sigma)$ with lifetimes ranging from $1~\mathrm{ps}$ to $\approx 10~\mathrm{ns}$. But among these particles, the highest mass occurs 
for the $\Lambda^0_b$ hadron or $B$ meson, having mass around $6~\mathrm{GeV}$~\cite{ pdg2}; hence the low $m_{DV}$ for the displaced vertex arising in the background is expected. 

The cross section and the event flow up to pre-selection for signals and backgrounds are noted in Table~\ref{uptopre}. Since the production channel for the signal is the DY pair-production, which only depends on the gauge coupling of the doubly charged scalar, the production cross section does not vary with the decay length of the particle, which is determined solely by the vacuum expectation value of the complex triplet for a given mass of the decaying scalar. 

\renewcommand{\arraystretch}{1.3}
\begin{table}
\begin{tabular}{|cc|c|c|c|c|c|}
\hline
\multicolumn{2}{|c|} {Benchmark Point} & cross section & Events & Events & Gen. & Pre-selection \\
Mass &  $c\tau$ &  [fb] & @HL-LHC & generated & selection & \\
\cline{5-6}
\hline
120~GeV & 5~mm & 470.95 & 1,412,850 & 1,000,000 & 572,826 & 421,885\\

$y_3 > y_2, y_1=0$ & & & & & & \\

120~GeV & 5~mm & 470.95 & 1,412,850 & 1,000,000 & 682,076 & 502,345 \\

$y_2 > y_1, y_3=0$ & & & & & & \\
\hline

300~GeV & 10~mm & 14.79 & 44,370 & 45,000 & 44,278 & 41,056\\
 & 10~cm  & & & & 44,787 & 41,158 \\
\hline

500~GeV & 10~mm & 1.60 & 4,800 & 10,000 & 9,860 & 9,456\\
  & 10~cm  & & & & 9,956  &  9,489\\
\hline

800 GeV & 10 mm & 0.15 & 450 & 10,000 & 9,959 & 9,575\\
  & 10 cm  & & & & 9,959 & 9,588  \\
\hline
\hline

QCD & & $7.18\times10^{11}$ & $2.15\times10^{15}$ & $2\times10^8$ & 12,583,104 &  759,969 \\
[1.0ex]

\hline
$t\bar{t}$ & & $5.81\times10^{5}$ & $1.74\times10^9$& $10^6$ & 354,256 & 321,317  \\
[1.0ex]

\hline
$V+\text{jets}$ & & $7.41\times10^{7}$ & $2.22\times10^{11}$ & $2\times10^6$ & 144,633 & 53,616 \\
$[V=W,Z]$ & & & & & & \\
[1.0ex]
\hline
\end{tabular}
\caption{Number of events up to pre-selection for signals and backgrounds at $14~ \text{TeV}$ $pp$ collider. For the definition of pre-selection, see fig.~\ref{cutflow}}
\label{uptopre}
\end{table}

\renewcommand{\arraystretch}{1.0}
\begin{table}
\centering
\begin{tabular}{|cc|c|c|c|c|}
\hline
\multicolumn{2}{|c|}{Benchmark Point} & BR1  & BR2 & BR3 & BR4\\

 Mass & $c\tau$ &  & & &\\
 \hline
 & & & & &  \\
 $120~\text{GeV}$ & 5~mm & 55.8\% & 43.6\% & $< 1\%$ & 0.0\%  \\
 $(y_3 > y_2, y_1=0)$ &  & & & & \\

 $120~\text{GeV}$ & 5~mm & 55.7\% & 0.0\% & 22.5 \% & 21.8\%   \\
 $(y_2 > y_1, y_3=0)$ &  &  & & &\\

 \hline
\end{tabular}
\caption{Branching ratios for the low-mass benchmark point, where, \\BR1=$BR(H^{\pm\pm} \rightarrow W^\pm W^{\star\pm})$, BR2=$BR(H^{\pm\pm} \rightarrow \tau^\pm \tau^{\pm})$, BR3=$BR(H^{\pm\pm} \rightarrow \mu^\pm \mu^{\pm})$, and BR4=$BR(H^{\pm\pm} \rightarrow e^\pm e^{\pm})$.}
\label{yukbp}
\end{table}

\renewcommand{\arraystretch}{1.3}
\begin{table}
\begin{tabular}{|cc|c|ccc|c|}
\hline
\multicolumn{2}{|c|} {Benchmark Point} & Criterion  & \multicolumn{3}{|c|} {Number of events} & Yield \\
\cline{4-6}
 Mass & $c\tau$ & & Hadronic & Semi-leptonic &  Leptonic & @HL-LHC \\
 & & & & & & \\
 \cline{1-7}

 $120~\text{GeV}$ & 5~mm & \textit{Cut-1} & 200,679 & 3,930 &  12,515 & 306,785  \\
 $(y_3 > y_2, y_1=0)$ & & \textit{Cut-2} & 186,527 & 3,930 &  8,729 & 281,439 \\
 &  & \textit{Cut-3} & 179,113 & 3,779 & 8,729 & 270,750  \\
  &  & \textit{Cut-4} & 174,632 & 3,636 &  8,425  &  263,787\\
 &  & \textit{Cut-5} & 106,191 & 2,454 & 7,769 & 164,487\\
 &  & & & & &  \\
 
 $120~\text{GeV}$ & 5~mm & \textit{Cut-1} & 198,944 & 8,795 &  12,420 & 310,970\\
 $(y_2 > y_1, y_3=0)$ &  & \textit{Cut-2} & 131,595 & 8,795 & 11,259 & 214,193\\
  &  & \textit{Cut-3} & 125,256 & 8,389 & 11,259 & 204,729\\
  &  & \textit{Cut-4} & 114,632  & 7,704 & 11,120 & 1,88,518\\
  &  & \textit{Cut-5} & 106,350 & 6,499 &  10,590 &  1,74,368 \\
 
\hline
\end{tabular}
\caption{The cut-flow table for signals at the $14~\text{TeV}$ $pp$ collider when the complex triplet couples to gauge bosons as well as leptons. For the definitions of \textit{Cut-1} to \textit{Cut-5} see fig.~\ref{cutflow}.}
\label{cutflowyuk}
\end{table}

\renewcommand{\arraystretch}{1.0}
\begin{table}
\centering
\begin{tabular}{|cc|c|ccc|c|}
\hline
\multicolumn{2}{|c|} {Benchmark Point} & Criterion  & \multicolumn{3}{|c|} {Number of events} & Yield \\
\cline{4-6}
 & & & Hadronic & Semi-leptonic &  Leptonic & @HL-LHC \\
 & & & & & & \\
 \cline{1-7}
 $300~\text{GeV}$ & 10~mm & \textit{Cut-1} & 35191 & 1976 & 1571 & 38195\\
 &  & \textit{Cut-2} & 32972 & 1976 & 1343 & 35782\\
  &  & \textit{Cut-3} & 32840 & 1971 & 1343 & 35647\\
  &  & \textit{Cut-4} & 32530 & 1945 & 1277 & 35251\\
  &  & \textit{Cut-5} & 30407 & 1904 &  1257 &   33098\\
  &  & & & & &  \\
  
 $300~\text{GeV}$ & 10~cm & \textit{Cut-1} & 34817 & 2002 & 1552 &  37833 \\
  & & \textit{Cut-2} & 32640 & 2001 & 1305 & 35442\\
 &  & \textit{Cut-3} & 32576 & 1993 & 1305 & 35371\\
  &  & \textit{Cut-4} & 27533  & 1499 & 868  & 29481\\
 &  & \textit{Cut-5} & 25498 & 1471 & 854 &  27433\\
 
 &  & & & & & \\

 $500~\text{GeV}$ & 10~mm & \textit{Cut-1} & 8219 & 653 & 395  & 4448 \\
 & & \textit{Cut-2} & 7520 & 653 & 334 & 4083\\
 &  & \textit{Cut-3} & 7370 &  649 & 334  & 4009\\
  &  & \textit{Cut-4} & 7246 & 631  & 321 &  3935\\
 &  & \textit{Cut-5} & 5647 &  597 &  320 &  3150\\

 &  & &  & & & \\
 
 $500~\text{GeV}$ & 10~cm & \textit{Cut-1} & 8164  &  673 & 398 & 4432\\
  &  & \textit{Cut-2} & 7475 & 673 & 327 & 4068\\
  &  & \textit{Cut-3} & 7440 & 661 & 327 & 4045\\
  &  & \textit{Cut-4} & 6520 & 524 & 246 &  3499\\
  &  & \textit{Cut-5} & 5962 & 503 &  246 & 3221\\
  &  & & & & & \\

  $800~\text{GeV}$ & 10~mm & \textit{Cut-1} & 7960   & 717 & 442  & 402 \\
  & & \textit{Cut-2} & 7157 & 717 & 306 & 360\\
 &  & \textit{Cut-3} & 7102 & 714 & 306 &  358\\
  &  & \textit{Cut-4} & 7005  & 700 & 300 & 353\\
 &  & \textit{Cut-5} & 6387 & 663 & 300 & 324\\

 &  & & & & & \\
 
 $800~\text{GeV}$ & 10~cm & \textit{Cut-1} & 8010   &  776  & 430 & 406 \\
 &  & \textit{Cut-2} & 7179  & 775 & 348 & 366 \\
  &  & \textit{Cut-3} & 7120 & 762 & 348 & 362 \\
  &  & \textit{Cut-4} & 6390 & 661 & 292 & 323\\
  &  & \textit{Cut-5} & 5939 & 636 & 292 & 302 \\
  &  & & & & & \\

\hline
\end{tabular}
\caption{Cut-flow for the signal when $H^{\pm\pm}$ decays exclusively into $W^\pm W^\pm$. For the definitions of \textit{Cut-1} to \textit{Cut-5}, please refer to fig.~\ref{cutflow}.}
\label{cutflownoyuk}
\end{table}

\renewcommand{\arraystretch}{1.05}
\begin{table}
\centering
\begin{tabular}{|cc|c|ccc|c|}
\hline
\multicolumn{2}{|c|} {Background} & Criterion  & \multicolumn{3}{|c|} {Number of events} & Yield \\
\cline{4-6}
 & & & Hadronic & Semi-leptonic &  Leptonic & @HL-LHC \\
 & & & & & & \\
 \cline{1-7}
 QCD & & \textit{Cut-1} & 70,846 & 0 & 0 & $7.6 \times 10^{11}$\\
  &  & \textit{Cut-2} & 67,174 & 0 & 0 & $7.2 \times 10^{11}$\\
  &  & \textit{Cut-3} &  43,690 & 0 & 0 & $4.7 \times 10^{11}$\\
  &  & \textit{Cut-4} & 40,340 & 0 & 0 & $4.3 \times 10^{11}$\\
  &  & \textit{Cut-5} & 0 & 0 & 0 & 0 \\
  &  & & & & &  \\
  
 $t\bar{t}$ & & \textit{Cut-1} & 258606 & 1 & 0 & $4.5\times10^8$ \\
 & & \textit{Cut-2} & 248144 & 0 & 0 & $4.3\times10^8$ \\
 &  & \textit{Cut-3} & 184482 & 0 & 0 & $3.2\times10^8$ \\
  &  & \textit{Cut-4} & 182388 & 0 & 0 & $3.1\times10^8$\\
 &  & \textit{Cut-5} & 0 & 0 & 0 & 0 \\
 
 &  & & & & & \\

 $V+\text{jets}$ & & \textit{Cut-1} & 16613 & 0 & 0 & $1.8\times10^9$ \\
 $[V=W,Z]$ & & \textit{Cut-2} & 15209 & 0 & 0 & $1.7\times10^9$ \\
 &  & \textit{Cut-3} & 9267 & 0 & 0 & $1.1\times10^9$\\
  &  & \textit{Cut-4} & 9166 & 0 & 0 & $1.0\times10^9$\\
 &  & \textit{Cut-5} & 0 & 0 & 0 & 0 \\

\hline
\end{tabular}
\caption{Cut-flow for backgrounds at $\sqrt{s}=14~\text{TeV}$ $pp$ collider. For the definitions of \textit{Cut-1} to \textit{Cut-5}, please refer to fig.~\ref{cutflow}.}
\label{cutflowbkgd}
\end{table}

\newpage
In the rest of this section, we will discuss the cut-flow in terms of events for different benchmark points.
\begin{itemize}
    \item {\textbf{$y_{\ell\ell} \ne 0$}}
    
    We consider a benchmark point with $m_{H^{\pm\pm}}=120~\mathrm{GeV}$, for which the corresponding cut-flow is presented in Table~\ref{cutflowyuk}. Assuming a diagonal triplet Yukawa matrix, with $Y_{ii}=y_i$, normal hierarchy implies $y_3>y_2>y_1$, while inverted hierarchy gives $y_2>y_1>y_3$. The branching ratios corresponding to this benchmark for both hierarchies are listed in Table~\ref{yukbp}. 
    
    \item {\textbf{$y_{\ell\ell} = 0$}}

    For negligible Yukawa coupling of the triplet scalar in the low-triplet-vev region, $H^{\pm\pm}$ emerges as an LLP. In this work, we have considered benchmark points above the $WW$ threshold for the $y_{\ell\ell} = 0 $ region. The analysis has been performed for three different masses of $H^{\pm\pm}$, and the corresponding cut-flow table is given in Table~\ref{cutflownoyuk}.

\end{itemize}

Since, in the first case, the doubly charged scalar couples to leptons, the leptonic channel has consistently higher efficiency than the semi-leptonic channel; on the other hand, for the second scenario, since the leptons are produced as the decay products of $W$ bosons, the efficiency of the leptonic analysis remains lower due to the suppression in the branching ratio. Finally, Table~\ref{cutflowbkgd} shows the cut-flow for the backgrounds. Since the required number of background MC samples could not be produced, we are not discussing the potential of a displaced lepton search in this context, even though the requirement of two displaced leptons can kill the background significantly, as evident from Table~\ref{cutflowbkgd}. Rather, we have discussed a displaced vertex search combining hadronic, semi-leptonic and leptonic channels in order to maximize the signal efficiency. For the hadronic analysis, the cut on $m_{DV}$ completely kills the backgrounds, which is also expected, as has been discussed previously in this section.

While our current strategy adopts a generalized displaced-vertex framework, we note that the doubly electrically charged nature of $H^{\pm\pm}$ offers powerful handles that experimental collaborations can exploit to further suppress backgrounds. Specifically, for macroscopic $c\tau$, $H^{\pm\pm}$ leaves a track with high $dE/dx$ in the inner tracker before decaying, which can be used to suppress backgrounds. It is also possible to do a charge reconstruction of the displaced vertex and to require the charge to be $\pm 2$. Additionally, one can also require two such displaced vertices from the pair-produced scalars. All these requirements can provide additional background rejection, and these can become a compelling direction for future experimental optimization in high pile-up scenarios.

\section{Results}
\label{result}

After applying the complete event selection criterion, no background event survived in our simulated samples, as shown in Table~\ref{cutflowbkgd}. Since the available Monte Carlo statistics are finite, this cannot be interpreted as evidence that the true expected background is exactly zero. Assuming that the number of background events follows a Poisson distribution, the observation of zero events corresponds to a one-sided 95\% confidence upper limit of 3 events on the true background expectation. 

While our simulated background samples yield zero surviving events after applying the invariant mass cut of $m_{DV} \ge 20~\mathrm{GeV}$, we acknowledge that finite Monte Carlo statistics cannot fully populate the rare tails of combinatorial background distributions. In real data, instrumental fakes, multi-vertex overlapping, and random track-crossing combinatorics can generate high-mass fake vertices that are not completely captured in this study. Our projected limits serve as a clean phenomenological baseline assuming an experimentally manageable background. In a real experimental search, backgrounds in this tail would be estimated using data-driven control regions or tail-extrapolation fits.

In order to derive the upper limit on the pair production cross section we have combined the three analysis
channels discussed in section~\ref{col}. Upper limit on the production cross section corresponds to $Z=1.645$ that gives the $5\%$ tail assuming an one-sided Gaussian distribution. The $Z-\mathrm{score}$ is computed using the Asimov formula given in ref-\cite{cowan}. Here we quote the upper limit assuming two values for the background, i.e $B=3 , 100$

\begin{figure}[htbp!]
	\centering
    \includegraphics[width=0.65\textwidth]{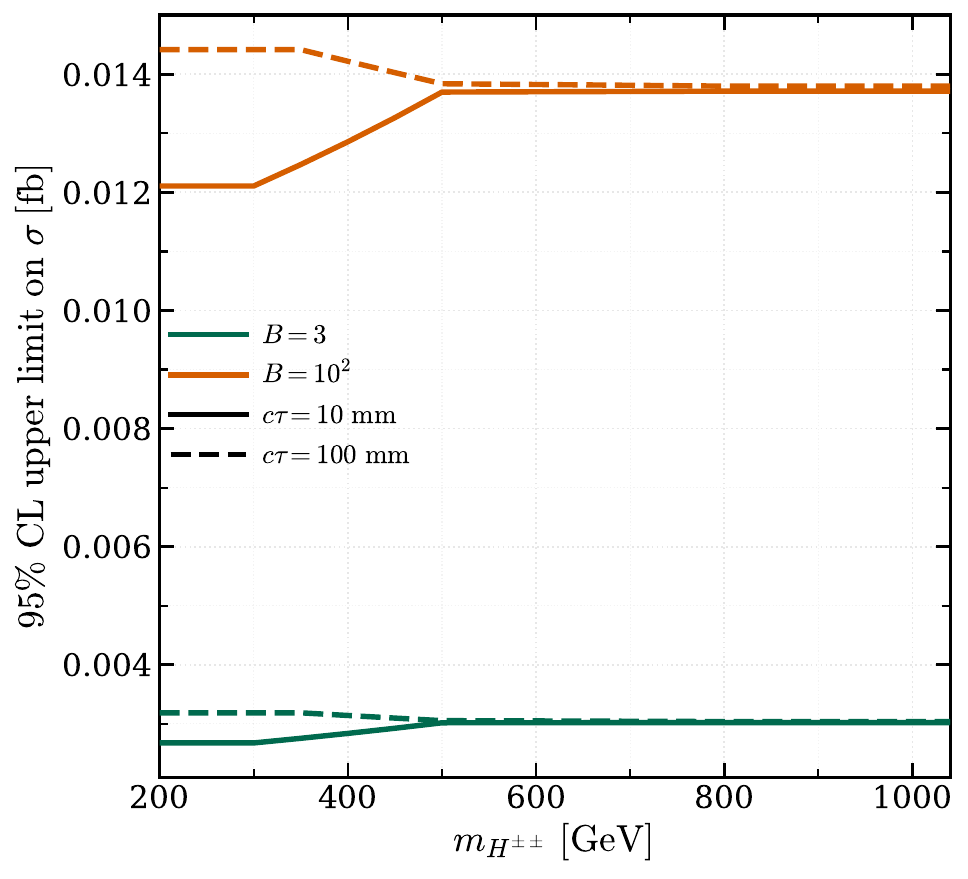}
	
	\caption{Expected upper limits at 95\% CL on the DY pair-production cross section of the doubly charged scalar when it couples to $W$ bosons only.} 
	\label{res_lim}
\end{figure}

\begin{figure}[h!]
	{\label{excl_prpx}\includegraphics[width=\textwidth]{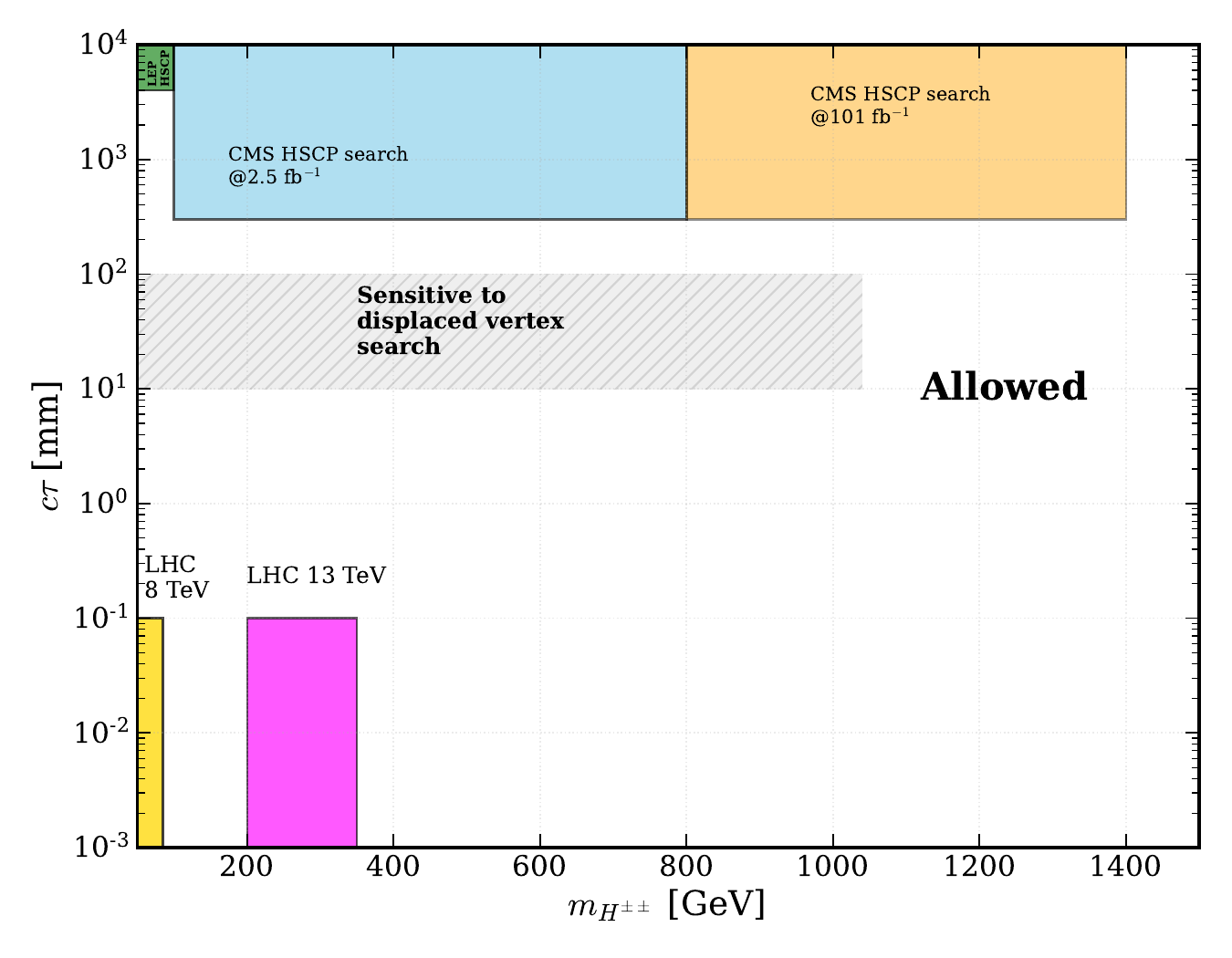}}
    
    \caption{Region of parameter space in the $m_{H^{\pm\pm}} - c\tau$ plane that has been explored in experiments to date when the doubly charged scalar couples predominantly to gauge bosons. The yellow region is excluded via a recast of the anomalous like-sign dilepton search by ATLAS in Run-1~\cite{ATLAS_2014, recast}; the magenta region is excluded via~\cite{atlas_2021}; the green region is excluded by~\cite{DELPHI_HSCP}; sky-blue region is excluded 
    by~\cite{cms_2016} and the orange region is excluded 
    by~\cite{cms_dedx_2025}. The region sensitive to the displaced vertex search that has been proposed here is shaded in gray color.}
	\label{excl_reg}
\end{figure}

\begin{itemize}
    \item $y_{\ell\ell} \ne 0:$
    
With efficiencies quoted in table-\ref{cutflowyuk}, table-\ref{uplim_yuk} shows the upper limit for $m_{H^{\pm\pm}}=120~\mathrm{GeV}$ and $c\tau=5~\mathrm{mm}$, for both the normal and the inverted hierarchies corresponding to a luminosity of $3000~\mathrm{fb}^{-1}$.

\renewcommand{\arraystretch}{1.3}
\begin{table}
\centering
\begin{tabular}{|cc|c|c|}
\hline
\multicolumn{2}{|c|} {Benchmark Point} & Background  & Upper limit on cross section [fb] \\

 Mass & $c\tau$ & & \\
 \cline{1-4}
 $120~\text{GeV}$ & 5~mm & 3 & 0.017  \\
 $(y_3 > y_2, y_1=0)$ & & 100 & 0.077 \\
 $120~\text{GeV}$ & 5~mm & 3 & 0.016  \\
 $(y_2 > y_1, y_3=0)$ & & 100 & 0.076 \\
\hline
\end{tabular}
\caption{Upper limit on cross section at 95\% CL corresponding to $\int \mathcal{L}dt = 3000~\mathrm{fb}^{-1}$ for $m_{H^{\pm\pm}}=120~\mathrm{GeV}$ and $c\tau = 5~\mathrm{mm}$ when $H^{\pm\pm}$ couples to both leptons and gauge bosons.}
\label{uplim_yuk}
\end{table}

As shown in table-\ref{uplim_yuk} the upper limit is almost 3 orders of magnitude smaller than the theoretically predicted cross section of  $470~\mathrm{fb}$ at $\mathcal{L}dt = 3000~\mathrm{fb}^{-1}$. Since, the upper limit on cross section is inversely proportional to the luminosity, this indicates that even though lowering the luminosity to $300~\mathrm{fb}^{-1}$ will increase the upper limit by around an order of magnitude, it will still be stringent enough to rule the scenario out at that luminosity.

    \item $y_{\ell\ell} = 0: $
    
    With the signal yields given in table~\ref{cutflownoyuk}, corresponding to
$\int \mathcal{L}\,dt = 3000~\mathrm{fb}^{-1}$, we derive the expected upper
limits on the pair-production cross section by combining the three analysis
channels discussed in section~\ref{col}. Figure~\ref{res_lim} shows the
resulting $95\%$~CL upper limits on $\sigma(pp \to H^{\pm\pm}H^{\mp\mp})$ for
$200~\mathrm{GeV} \le m_{H^{\pm\pm}} \le 1.05~\mathrm{TeV}$ and for
$c\tau = 10~\mathrm{mm}$ and $100~\mathrm{mm}$. It is important to note that though we have performed the simulation for $m_{H^{\pm\pm}} = 300~\mathrm{GeV},500~\mathrm{GeV},800~\mathrm{GeV}$, we have linearly interpolated the efficiencies for the other masses between $300~\mathrm{GeV}-800~\mathrm{GeV}$ and extrapolated to $200~\mathrm{GeV}$ and $1.05~\mathrm{TeV}$ in the lower and upper end respectively as shown in fig-\ref{res_lim}.

Below about $500~\mathrm{GeV}$ the limit is marginally weaker for
$c\tau = 100~\mathrm{mm}$ than for $c\tau = 10~\mathrm{mm}$. This reflects the
material-map veto: displaced vertices reconstructed beyond the outermost layer
of the material map of ref.~\cite{cms_matmap}, at a radial distance of
$150~\mathrm{mm}$, are rejected. At $c\tau = 100~\mathrm{mm}$ a sizeable
fraction of decays occur outside this radius, and this fraction is largest for
the lighter scalars, which are produced with the largest boost. As the mass
increases the boost decreases and these vertices move back inside the fiducial
volume, while for $c\tau = 10~\mathrm{mm}$ the same reduction in boost shifts
the decays into the dense inner-detector material, where they are likewise
vetoed.  Also for $c\tau = 10mm$, smaller boosts will enhance the population of vertices with smaller displacement affecting the reconstruction efficiency as well. The two limits therefore converge above $\sim 500~\mathrm{GeV}$.

    Comparing these limits with the corresponding theoretical pair-production cross sections, we find the displaced vertex search is sensitive to the full considered mass range for $c\tau\in(10~\mathrm{mm},100~\mathrm{mm})$. Fig.~\ref{excl_reg} shows the resulting sensitive region in the $m_{H^{\pm\pm}}$-$c\tau$ plane.
\end{itemize}

\section{Summary and conclusion}
\label{conc}

In this work, we have identified the unexplored regions in the $m_{H^{\pm\pm}}-c\tau$ plane that motivate dedicated searches for long-lived doubly charged scalars. We have considered two scenarios: one in which the doubly charged scalar couples to leptons and another in which it is fermiophobic. For each scenario, we have identified the regions that remain largely unexplored by existing searches and investigated the prospects of probing them through displaced-vertex signatures.

To maximize the signal sensitivity, we have considered three complementary final-state topologies, namely the hadronic, semi-leptonic, and leptonic channels. We demonstrate that the proposed displaced-vertex search can probe doubly charged scalars with masses in the range $200~\mathrm{GeV}-1~\mathrm{TeV}$ and decay length in range $c\tau=10~\mathrm{mm}-100~\mathrm{mm}$ at the HL-LHC when doubly charged scalar couples only to gauge bosons. On the other hand, in the presence of the Yukawa coupling of the triplet scalar we have derived the upper limit on cross section for a $H^{\pm\pm}$ with a mass of $120~\mathrm{GeV}$ and $c\tau = 5~\mathrm{mm}$, which further indicates that such a scalar could potentially be ruled out with an integrated luminosity of $300~\mathrm{fb}^{-1}$. Since the proposed strategy does not rely on a specific final state, but instead enhances the sensitivity through the combination of complementary final states, it provides a relatively model-independent framework applicable to doubly charged scalars arising in a variety of scenarios, including the Type-2 seesaw, composite Higgs, and VLF models.

Overall, our results demonstrate the importance of displaced-vertex searches in exploring the long-lived particle parameter space of doubly charged scalars. Their decays into same-sign $W$ bosons and leptons can give rise to distinctive experimental signatures that are complementary to conventional prompt searches. The combined analysis of hadronic, semi-leptonic, and leptonic final states significantly enhances the sensitivity and provides a broader probe of the relevant parameter space. 

Future studies incorporating a more detailed detector simulation, including \textsc{Geant4}-based modeling of instrumental backgrounds and pile-up effects, together with data-driven background estimates, will be important for further improving the robustness of the proposed search. In addition, optimized charge-sensitive observables, such as $dE/dx$ and vertex-charge information, could provide further discrimination and strengthen the prospects for discovering or constraining long-lived doubly charged scalars at the HL-LHC and future colliders.
\section{Acknowledgment}

The authors thank S. Mrenna for useful suggestions on the simulation of long-lived particles using \textsc{Pythia8}. The authors also thank Biswarup Mukhopadhyaya for useful discussions. The work of RG has been supported by the Institute Post-doctoral Fellowship provided by Indian Institute of Technology (IIT) Kanpur. NKP would like to acknowledge IIT Kanpur for providing financial support through PhD fellowship. The work of SM is supported by an initiation grant ($\text{IITK}/\text{PHY}/2023282$) received from IIT Kanpur. The work of BB is supported by the Core Research Grant CRG/2022/001922 of the Anusandhan National Research Foundation (erstwhile Science and Engineering Research Board (SERB) scheme), Government of India.

\end{document}